\documentclass[aps,physrev,preprint,superscriptaddress]{revtex4-2}

\usepackage{graphicx}
\usepackage{amsmath,amssymb,bm}
\usepackage{siunitx}
\usepackage{hyperref}
\usepackage{subcaption}
\usepackage{rotating}   
\usepackage[percent]{overpic}

\begin{document}

\title{A Unified Spectrum for Turbulence in Microfluidic Flow}

\author{Chit Yau Kuan}
\affiliation{Department of Biomedical Engineering, The Chinese University of Hong Kong, Shatin, Hong Kong SAR, China}

\author{Xiaochen Liu}
\affiliation{School of Biomedical Engineering, Faculty of Engineering, The University of Sydney, Sydney, NSW, 2008, Australia}

\author{Yi Ping Ho}
\affiliation{Department of Biomedical Engineering, The Chinese University of Hong Kong, Shatin, Hong Kong SAR, China}

\author{Ken-Tye Yong}
\affiliation{School of Biomedical Engineering, Faculty of Engineering, The University of Sydney, Sydney, NSW, 2008, Australia}


\begin{abstract}
Turbulent-like flows in microfluidic systems arise through non-inertial
mechanisms---elastic polymer stress, bacterial self-propulsion, electrokinetic
body forces, and interfacial tension gradients---each operating at Reynolds
numbers far below the classical inertial turbulence threshold.
The resulting energy spectra follow power-law scaling
$E(k)\sim k^{-m}$ with exponents that deviate systematically from the
Kolmogorov $k^{-5/3}$ law, yet no common framework
relates these exponents to the underlying physics.
We extend the Kolmogorov--Pao spectral model to microfluidic turbulence
through a master spectrum governed by an adaptive slope exponent $\sigma$,
expressed as a function of regime-specific dimensionless groups---the
electric Rayleigh number $Ra_e$ for electrokinetic flows, the Capillary
number $Ca$ for interfacial flows, the Weissenberg number $Wi$ for elastic
flows, and the Reynolds number $Re$ for active suspensions---together with
two spectral cutoffs: the classical viscous wavenumber $k_\eta$ and a
physics-specific wavenumber $k^\star$ defined by the timescale equality
$\tau_\mathrm{regime}=\tau_\mathrm{viscous}$.
The master spectrum is calibrated against published spectra: the $k^{-5/3}$ to $k^{-7/5}$ cascade
transition in electrokinetic turbulence~\cite{Shi2025_quadCascadeEKT},
the $k^{-4.5}$ interfacial spectrum~\cite{Padhan2024_interfaceTurbulence},
the $k^{-8/3}$ active-suspension spectrum~\cite{Wensink2012MesoScaleTurbulence},
and the $f^{-3.5}$ elastic turbulence spectrum~\cite{Groisman2000_elasticTurbulence},
spanning Reynolds numbers from $Re\sim10^{-4}$ to $Re\sim1$.
The effective spectral slope $-5/3-\sigma$ constitutes a spectral fingerprint
that identifies the dominant non-inertial mechanism directly from a measured
spectrum, providing a compact design tool that complements, rather than merely reproduces, computationally
expensive direct numerical simulation.
\end{abstract}


\maketitle

\section{Introduction}

Turbulence in microfluidic systems presents a fundamental challenge to classical theory: whereas macroscale turbulence arises from inertial instabilities at $\mathrm{Re} > 2000$, a growing body of evidence demonstrates that turbulent-like states emerge in microfluidic flows ($\mathrm{Re} < 100$) through mechanisms that operate entirely outside the inertial framework---including elastic polymer stress~\cite{Groisman2000_elasticTurbulence}, active bacterial locomotion~\cite{Wensink2012MesoScaleTurbulence}, electric body forces~\cite{Wang2016_microEKT}, geometric and compliance-induced instabilities~\cite{Kumaran2016_softWallTurbulence}, interfacial tension gradients~\cite{Padhan2024_interfaceTurbulence}, and thermodynamic criticality~\cite{Hurtan2025_supercriticalMiniaturization}. Each mechanism generates and sustains flow instability through a distinct physical pathway, producing energy spectra that deviate characteristically from the Kolmogorov $k^{-5/3}$ inertial-range law, while maintaining the scale-invariant power-law behavior ($E(k)\sim k^{-m}, \quad k \in [k_1,k_2],\quad m \in \mathbb{R}$).
Understanding and predicting these spectra is essential for microfluidic applications where turbulence governs mixing efficiency and energy transport, including protein folding kinetics~\cite{Roder2004RapidMixing} and cell isolation~\cite{Che2016_vortexCTC}.

Despite this diversity of turbulent regimes, existing spectral models for microfluidic systems remain predominantly empirical---constructed by fitting Kolmogorov's inertial-range scaling to experimental data without accounting for the distinct physics of each regime. Classical Kolmogorov--Pao theory~\cite{Kolmogorov1941_localStructure,Pao1965_spectrum} assumes turbulence is sustained by fluid inertia, with energy cascading from large injection scales to a viscous dissipation cutoff---assumptions that break down when the driving mechanisms are non-inertial forces such as elastic stress, active forcing, or external field coupling. As a result, no single dimensionless parameter or governing equation currently unifies turbulence onset prediction or spectral characterization across microfluidic regimes, and the \textit{de facto} approach of full direct numerical simulation or CFD~\cite{Ebner2023_CFDmicrofluidics} remains prohibitively expensive for systematic design exploration.

Here we address this gap by proposing a predictive master spectrum for microfluidic turbulence, built upon the Kolmogorov--Pao framework and extended to incorporate system-specific physics. The spectral slope is treated as an adaptive parameter governed by key dimensionless groups---the Reynolds number, Mach number, Weissenberg number, activity parameter, and field-coupling parameters such as electric Rayleigh number ($Ra_e$) for electrokinetic and Capillary number ($Ca$) for interfacial flows---allowing the model to reproduce the distinct slopes observed across regimes. 
Two spectral cutoffs generalize the single viscous cutoff of classical Kolmogorov-Pao theory. 
The first, a generalized viscous cutoff at $k_{\eta}$, accounts for the low-Reynolds-number 
character of microfluidic turbulence: viscous forces dominate across a wider range of scales 
than in classical inertial turbulence, compressing the cascade range and shifting viscous 
dissipation onset to larger scales. 
The second, a physics-specific cutoff at $k_{\star}$, captures the characteristic length scale 
imposed by each non-inertial driving mechanism—such as the electric Kolmogorov scale 
$\ell_{de}$ in electrokinetic turbulence~\cite{Wang2016_microEKT} or the swimmer body length 
$\ell$ in active suspensions~\cite{Wensink2012MesoScaleTurbulence} (Table~\ref{tab:physics_cutoffs})—below which the mechanism 
can no longer sustain energy input and the spectrum rolls off before viscosity acts. 
Together, $k_{\eta}$ and $k_{\star}$ bound the energy cascade in each regime. We demonstrate that this master spectrum reproduces reported energy 
spectra across known microfluidic turbulence regimes, providing a compact predictive tool 
for design-space exploration that bypasses full numerical simulation while retaining 
compatibility with DNS or experiment for detailed validation.

\section{Governing equations and methodology}
\subsection{Theoretical foundation of the master equation}

The starting point is the incompressible Navier--Stokes equations under the context of classical turbulence analysis with forcing $\mathbf{f}$:
\begin{equation}
\partial_t \mathbf{u} + (\mathbf{u} \cdot \nabla)\mathbf{u} 
= -\nabla p / \rho + \nu \nabla^2 \mathbf{u} + \mathbf{f}, 
\qquad \nabla \cdot \mathbf{u} = 0.
\label{eq:NS}
\end{equation}
Fourier transforming the velocity field and averaging over spherical wavenumber shells
defines the spectral energy density,
\begin{equation}
E(k,t) = \frac{1}{2} \left| \hat{\bm{u}}(\bm{k},t) \right|^2 ,
\label{eq:specdef}
\end{equation}
such that the total kinetic energy per unit mass is
\begin{equation}
\frac{1}{2} \langle |\bm{u}|^2 \rangle = \int_0^{\infty} E(k,t)\, dk .
\label{eq:Etot}
\end{equation}

The Fourier transform of the Navier--Stokes equations reformulates the real-space, time-domain momentum balance into a wavenumber-space representation. Multiplying the resulting equation by the complex conjugate of the velocity field and taking the real part converts the momentum equation into an evolution equation for the spectral kinetic energy [Eq.~\eqref{eq:specdef}]. Upon averaging over spherical shells in wavenumber space---assuming statistical isotropy---this procedure yields the classical Lin equation governing the spectral energy budget (see Refs.~\cite{Owen1956_reviewLinHydrodynamicStability,pope2000} for derivation):
\begin{equation}
\partial_t E(k,t) = T(k,t) + F(k,t) - 2 \nu k^2 E(k,t) ,
\label{eq:Lin}
\end{equation}
where $F(k,t)$ is the spectral forcing (energy injection at large scales),
$T(k,t)$ is the nonlinear transfer (advective redistribution across scales, conserving total energy),
and $2\nu k^2 E(k,t)$ is viscous dissipation.

In a statistically steady state, the mean spectral distribution is stationary:
\begin{equation}
\langle \partial_t E(k,t) \rangle = 0 ,
\label{eq:steady}
\end{equation}
so injection and dissipation balance globally. Within the inertial range  
($k_L \ll k \ll k_\eta$), large-scale forcing and small-scale viscous dissipation are negligible 
 [$F(k) \simeq 0$, $2\nu k^2 E(k) \simeq 0$], and the Lin equation reduces to $T(k) \simeq 0$. This condition does not imply the absence of energy transfer; rather, it states that no energy accumulates at any individual scale. The cascade itself is carried by the cumulative flux
\begin{equation}
\Pi(k) = \int_k^\infty T(k') \, dk' .
\label{eq:flux}
\end{equation}
which represents the rate at which energy is transmitted through wavenumber \(k\) toward smaller scales. The condition \(T(k)\simeq 0\) is equivalent to
$\partial \Pi/\partial k\simeq 0,$
meaning that the flux is independent of \(k\) across the inertial range. Energy therefore passes through each scale at a constant rate without local deposition or removal.
Because this constant flux persists throughout the cascade and is ultimately set by the rate at which energy is dissipated at the smallest scales, it equals the mean dissipation rate $\varepsilon$:

\begin{equation}
\Pi(k) = \varepsilon .
\label{eq:eps}
\end{equation}
This constant-flux condition is the central statement of the Kolmogorov cascade.

In microscale and microfluidic flows, however, the premise of scale separation--namely that energy injection and dissipation are confined to large and smallest scale respectively--is generally violated. Additional power transfer and removal can possibly occur across the broad inertial range of wavenumber through a non-inertial mechanism, including scalar forcing (electrokinetic driving ~\cite{Wang2016_microEKT}), activity (micro-suspension motion ~\cite{Thampi2016_activeMicromachines}), interfacial stress ~\cite{Padhan2024_interfaceTurbulence}, and compressive or shock-related transfer~\cite{White2019_supersonicPlasma}. 
These processes introduce scale-dependent sources and sinks of kinetic energy that coexist with inertial transfer within the nominal inertial range. Formally, they enter the Lin equation as additional contributions to the right-hand side that are distributed across \(k\) rather than localized at the injection and dissipation scales. Consequently, $\partial \Pi/\partial k$
no longer vanishes within the inertial range, and the assumption of a constant energy flux --- together with the associated Kolmogorov \(k^{-5/3}\) scaling --- breaks down. The effective energy throughput reflects a cumulative balance between inertial transfer and additional scale-dependent exchange channels. 

\begin{equation}
\Pi_{\mathrm{tot}}(k)
=
\Pi_{\mathrm{inertial}}(k)
+\Pi_{\mathrm{scalar}}(k)
+\Pi_{\mathrm{stress}}(k)
+\Pi_{\mathrm{compressive}}(k)
\label{eq:flux_decomposition}
\end{equation}

The proposed renormalization of the flux balance accounts for the modified energy budget and provides a consistent extension of the classical cascade framework to microfluidic turbulence-like regimes.



By dimensional arguments, this determines the inertial-range spectrum in the microfluidic turbulence as
\begin{equation}
E(k) = C_K \, \varepsilon^{2/3} k^{-5/3}\left(\frac{k}{k_0}\right)^{-\sigma}.
\label{eq:powerlaw}
\end{equation}

The model conserves the classical Kolmogorov equation with a self-similar power-law modulation to summarize the effect of total flux on the slope deviation. The secondary wavenumber \(k_0\) is defined as the inverse of a characteristic
crossover length \(\ell_0\),
\begin{equation}
\label{eq:k_0}
k_0 \equiv \frac{1}{\ell_0},
\qquad
\end{equation}
Physically, \(k_0\) marks the onset of cascade-like range which the modified spectral slope develops and the vortex sustains. \cite{Yerasi2024_preservingLargeScaleElasticTurbulence}. In the limit \(\sigma = 0\), the formulation reduces to the
classical Kolmogorov form.

For incompressible Newtonian fluids, the viscous dissipation rate is exactly

\begin{equation}
\varepsilon = 2\nu \left\langle s_{ij}s_{ij} \right\rangle,
\end{equation}

where \(s_{ij}\) is the rate-of-strain tensor \cite{pope2000}. This expression is valid across scales under the continuum assumption (\(Kn \ll 1\)). Under incompressibility, it can be rewritten as

\begin{equation}
\varepsilon
=
\nu
\left\langle
\left(
\frac{\partial u_i}{\partial x_j}
\right)^2
\right\rangle,
\end{equation}

which implies the scaling

\begin{equation}
\varepsilon
\sim
\nu \dot{\gamma}^{\,2},
\end{equation}

where the characteristic strain rate scales as $\dot{\gamma} \sim U/L$.

With the scale-dependent modulation on the energy flux profile, the effective slope under microfluidic regimes is adjusted by $\sigma$, a term that considers the non-inertial effect from Reynolds number ($\mathrm{Re}$), compressibility ($M$), activity/elasticity ($\chi$), normalized Weissenberg number ($Wi^{norm}$), Capillary number ($Ca$), and normalized electric Rayleigh number ($Ra_e^{norm}$) on the energy dissipation slope: 

\begin{equation}
\label{eq:slope}
\sigma \;=\; \; a_{1}\,\frac{M^{2}}{1+M^{2}}
\;+\; a_{2}\,\frac{\chi^{2}}{1+\chi^{2}}
\;+\; a_{3}\,\mathrm{Re}^{-1/4}
\;+\; f(Wi^{norm})
\;+\; f(Ca)
\;+\; f(Ra_e^{norm}).
\end{equation}

\begin{table}[h]
\centering
\caption{Dimensionless coefficients used in Eq.~\eqref{eq:slope}.}
\begin{tabular}{c c l c}
\hline\hline
\textbf{Coefficient} & \textbf{Approximated value} & \textbf{Physical interpretation} & \textbf{Reference} \\
\hline
$a_1$ & $1/3$   & Compressibility correction (Mach number) & ~\cite{White2019_supersonicPlasma, Aniskin2015_supersonicMicrojets} \\
$a_2$ & $-1/6$  & Activity/elasticity correction           & ~\cite{Thampi2016_activeMicromachines, Wensink2012MesoScaleTurbulence} \\
$a_3$ & $3/20$  & Micro-scale Reynolds correction          & ~\cite{Sofiadis2022MicropolarRe} \\
\hline\hline
\end{tabular}
\label{tab:coeffs}
\end{table}

\begin{table}[h]
\centering
\caption{Representative regimes and mechanisms defining the physics cutoff $k_\star$.}
\begin{tabular}{l l c}
\hline\hline
\textbf{Regime} & \textbf{Mechanism} & \textbf{Cutoff definition} \\
\hline
Compressible   & Shock width $\delta_s$                            & $k_\star = 1/\delta_s$ \\
Active matter  & Swimmer size $\ell_{swimmer}$                                 & $k_\star = 1/R_v$ \\
Viscoelastic   & Relaxation length $\ell_{\mathrm{relax}}$         & $k_\star = 1/\ell_{\mathrm{relax}}$ \\
Electrokinetic & Debye length $\lambda_D$                          & $k_\star = 1/\lambda_D$ \\
Soft-wall      & Wall scale $\delta_{\mathrm{wall}}$               & $k_\star = 1/\delta_{\mathrm{wall}}$ \\
\hline\hline
\end{tabular}
\label{tab:physics_cutoffs}
\end{table}

where the Reynolds number is defined as $\mathrm{Re} = {U L}/{\nu}$, with $U$ the characteristic velocity, $L$ the characteristic length scale, and $\nu$ the kinematic viscosity, and the Mach number is defined as $M = V/c$, with $V$ the fluid velocity and $c$ the speed of sound. Although the coefficients $a_i$ are not derived from first principles, they are not introduced as arbitrary fitting parameters. Instead, each coefficient is asymptotically anchored to the inertial-range scaling exponent reported in the regime where the corresponding physical mechanism is dominant. The bounded functional forms enforce a controlled and continuous crossover and ensure recovery of the Kolmogorov limit in the absence of that mechanism. In this way, the formulation is constrained by known asymptotic behaviors and physical limits rather than relying on unconstrained phenomenological interpolation.

Beyond the inertial range, viscosity dominates the irreversible conversion of kinetic 
energy into heat, thereby producing entropy. This entropy-generating sink necessitates 
inclusion of a viscous dissipation term in the spectral balance. Following Pao's closure, 
balancing nonlinear transfer and dissipation yields an exponential roll-off at the 
Kolmogorov wavenumber $k_\eta = (\varepsilon / \nu^3)^{1/4}$:
\begin{equation}
E(k) = C_K \, \varepsilon^{2/3} k^{-5/3} 
\exp\left[ -\frac{3}{2} \left( \frac{k}{k_\eta} \right)^{4/3} \right].
\label{eq:pao}
\end{equation}

This form accounts for the universal viscous cutoff. However, the assumptions underlying Pao's dissipation model do not hold in microfluidic turbulence. Microfluidic constraints (strong wall confinement, active micro-motion, electrokinetic forcing etc.) impose spatially varying dissipation, and compared to classical turbulence where inertial range and viscous cutoff are distinctly separated, viscosity dominates at microscale Reynolds number under micro-flow, so the inertial range is short and underdeveloped. We therefore generalize the  viscous prefactor $\gamma_v$ and $\alpha_{v}$, while ensuring recovery of the Kolmogorov--Pao form in the appropriate asymptotic limit ($\gamma_v=3/2$ and $\alpha_{v}=4/3$).

In addition to natural viscous 
dissipation, microfluidic systems exhibit system-specific 
entropy-producing mechanisms---including polymer relaxation, conductivity gradients, 
active stresses from bacteria or motors, wall compliance, and other forms of active forcing. 
These processes irreversibly degrade kinetic energy and act as additional sinks within the 
cascade, leading to premature dissipation prior to the viscous cutoff. To capture this, we 
introduce an additional physics-based damping term, yielding a generalized 
master spectrum that supplements the inertial power law with both viscous and 
physics-specific cutoff functions:
\begin{equation}
E(k) = k_0^{\sigma}\,C_K \, \varepsilon^{2/3} k^{-5/3-\sigma} 
\exp\left[ -\gamma_v \left( \frac{k}{k_\eta} \right)^{\alpha_v} \right]
\exp\left[ -\gamma_p \left( \frac{k}{k_\star} \right)^{\alpha_p} \right].
\label{eq:master}
\end{equation}

Here $k_\eta$ is the Kolmogorov wavenumber (viscous cutoff), the inertial-range scaling exponent $-5/3-\sigma$ defines an effective slope exponent, where 
$\sigma$ quantifies systematic deviations from the classical Kolmogorov cascade induced by non-inertial mechanisms, and $k_\star$ denotes a physics-specific cutoff with the following definition \cite{Wang2016_microEKT}:
\begin{equation}
\tau_{\mathrm{regime}}=\tau_{\mathrm{viscous}}
\label{eq:k_star}
\end{equation}
The timescale equality identifies the crossover wavenumber at which the viscous diffusion time ($\tau_{viscous}=\ell^{2}/v$~\cite{FouxonLebedev2003, Wang2016_microEKT}) equals the characteristic timescale of the regime-specific driving, marking the onset of viscous dominance over that mechanism.
 The decay exponents $\alpha_{v,p}$ and prefactors , $\gamma_{v,p}$ are not introduced as free fitting parameters; rather, they encode the scale dependence of the underlying dissipation mechanisms and are constrained by asymptotic consistency with the classical Kolmogorov–Pao viscous limit. In particular, the functional form of the damping terms ensures recovery of the universal inertial–viscous transition in the absence of additional physics, while allowing controlled deviations when non-inertial dissipation pathways are present. This compact expression unifies classical turbulence scaling 
with microfluidic-specific energy sinks, allowing an alternative prediction of spectral structure 
prior to computationally expensive CFD simulations.

\subsection{Constraints of the model}

The proposed master spectrum is intended for statistically stationary
microfluidic turbulence-like states and is not formulated to predict the
initial linear onset of instability from a laminar base flow. Rather, it
applies once a sustained broadband fluctuation state has already been
established. The formulation further assumes that the spectral response is
governed by a dominant non-inertial driving mechanism, or by a limited
combination of mechanisms whose effects can be represented through additive
corrections to the spectral slope and cutoff scales.

A second constraint is the existence of a finite scaling range over which the
spectrum carries physically meaningful information, including the cascade
extent, fluctuation intensity, energy level, and characteristic vortex size.
Accordingly, it is necessary to distinguish the cascade topology associated
with different non-inertial regimes.

The first class comprises single-pass forward-cascade regimes,
for which
\begin{equation}
k_0 \leq k \leq k_{\min}(k_\eta,k_\star),
\label{eq:forward_topology}
\end{equation}
and energy injected at the onset scale $k_0$ is transferred monotonically
toward higher wavenumber until dissipation at
$k_{\min}$. This topology is the natural analogue of a forward cascade within a finite
microfluidic spectral window similar in classical inertial turbulence.

The second class comprises bidirectional-cascade regimes, for which
the spectral transfer follows
\begin{equation}
k_0 \rightarrow k_m \rightarrow k_{\min},
\qquad
k_m < k_0 \leq k_{\min},
\label{eq:bidirectional_topology}
\end{equation}
where $k_m$ denotes the mesoscale turning wavenumber reached by the inverse
cascade. In this case, the primary energy injection occurs near the
microscopic forcing scale $k_0$---for example, the inverse body-length scale
of active swimmers---which lies close to the high-wavenumber range where
viscous dissipation also becomes important, i.e.,
$k_0 \approx k_\eta$ or $k_0 \approx k_\star$. The injected energy therefore
first undergoes an upscale transfer toward mesoscopic structures at $k_m$,
before a secondary forward transfer returns energy to higher wavenumber for
final viscous dissipation at $k_{\min}$.
\subsubsection{Conditional role of Kolmogorov cutoff}
When the time-averaged Reynolds number exceeds unity ($Re\geq 1$) across the spectral range, $k_\eta$ operates as an independent spectral cutoff and the Kolmogorov scale formula 

\begin{equation}
\eta = \left(\frac{\nu^{3}}{\varepsilon}\right)^{1/4},
k_\eta=\frac{1}{\eta}
\label{eq:Kolmogorov scale}
\end{equation}

gives the physically correct scale. When the time-averaged $Re\leq1$ throughout the spectrum, $k_\eta$ is not independently realised, and the sole observable cutoff is $k_\star$ with the reported spectral roll-off marked as a cross validation ($k_{\eta,spectrum}$). The master spectrum retains the $k_\eta$ exponential for formal consistency with the classical Kolmogorov-Pao limit while it does not affect the spectral slope $-5/3-\sigma$.

\section{Results}

\subsection{Proof-of-principle validation of the master equation}
\subsubsection{Microelectrokinetic turbulence}

Shi~\textit{et al.}~\cite{Shi2025_quadCascadeEKT} investigated electrokinetic turbulence across a range of electric Rayleigh numbers, $Ra_e$, in a Y-shaped PDMS micromixer of height $h = 100~\mu\mathrm{m}$, initial width $w_0 = 620~\mu\mathrm{m}$, and length $\ell = 9~\mathrm{mm}$. Two co-flowing streams with an electric conductivity ratio $\sigma_1:\sigma_2 = 1:5000$ were driven at $3~\mu\mathrm{L/min}$ (AC electric field at 130 kHz)each, giving a bulk Reynolds number
$Re = U w_0/ \nu = 1$.

Velocity fluctuations were measured by laser-induced fluorescence photobleaching anemometer (LIFPA) with spatial resolution $180~\mathrm{nm}$ and temporal resolution $\mathcal{O}(10~\mu\mathrm{s})$, enabling direct access to the inertial-range velocity spectrum $E_u(k)$.

The Taylor microscale and Taylor Reynolds number were extracted directly from the velocity time series via
\begin{equation}
    \lambda = \frac{u_{\mathrm{rms}}}
    {\langle (\partial u'/\partial x)^2 \rangle^{1/2}},
    \qquad
    Re_\lambda
    =
    \frac{u_{\mathrm{rms}}\lambda}{\nu}.
    \label{eq:Taylor microscale}
\end{equation}
At $Ra_e = 2.89\times10^6$, these are $\lambda = 2.01\times10^{-5}~\mathrm{m}$ and $Re_\lambda = 0.018$; at $Ra_e = 4.52\times10^6$, they decrease to $\lambda = 1.85\times10^{-5}~\mathrm{m}$ and $Re_\lambda = 0.016$. Using the root-mean-square velocity fluctuation to compute isotropic dissipation rate~\cite{taylor1935statistical,batchelor1953homogeneous}
\begin{equation}
\varepsilon
=
\frac{15 \nu u_{\mathrm{rms}}^2}{\lambda^2}.
\label{eq:isotropic dissipation}
\end{equation}
This yields $\varepsilon \approx 2.98\times10^{-2}~\mathrm{m^2\,s^{-3}}$ and $3.28\times10^{-2}~\mathrm{m^2\,s^{-3}}$ at the two $Ra_e$ conditions, respectively (Table~\ref{tab:summary}). 

The viscous cutoff wavenumber is estimated as the Kolmogorov scale (Eq. ~\ref{eq:Kolmogorov scale}) at $1.31\times10^4~\mathrm{m^{-1}}$ to $1.35\times10^4~\mathrm{m^{-1}}$ with respect to $Ra_e = 2.89\times10^6$ and $Ra_e = 4.52\times10^6$. The spectrum-observable viscous rolloff in the measured $E_u(k)$ from Shi~\textit{et al.} report at approximately
$k_{\eta,spectrum} \approx (1\text{--}2)\times10^4~\mathrm{m^{-1}}$ ($\pm 10\%$ compared to the Kolmogorov scale).

The central observation of Shi~\textit{et al.} is a systematic transition in the cascade slope of $E_u(k)$ as $Ra_e$ increases: at $Ra_e = 2.89\times10^6$, the velocity spectrum follows $-5/3$,
consistent with a constant-$\Pi_u$ subrange, where the kinetic energy flux is conserved; at $Ra_e = 4.52\times10^6$, the slope steepens to $-7/5$,
consistent with a constant-$\Pi_\sigma$ subrange. 

Within the master-spectrum framework [Eq.~\eqref{eq:master}], this slope
evolution is encoded in the correction $f(Ra_e^{\mathrm{norm}})$, where
$Ra_e^{\mathrm{norm}}$ is the symmetric, frequency-corrected electric Rayleigh
number defined in the Supplemental Material, which places the Shi~\textit{et al.}~\cite{Shi2025_quadCascadeEKT}
and Wang~\textit{et al.} ~\cite{Wang2016_microEKT} data on a common scale. Slope correction $f(Ra_e^{\mathrm{norm}})$ is constructed with the two asymptotic limits fixed by independent arguments while the transition between
them is fitted to the data: at weak forcing the electrokinetic body
force is negligible, the cascade reverts to the inertial constant-$\Pi_u$
subrange, and Kolmogorov recovery requires $f\to0$, i.e.\ a slope $-5/3$ under high $Re$. At
strong forcing the kinetic-energy flux is subjected to the scalar (conductivity), and
the quad-cascade analysis of Shi~\textit{et al.}~\cite{Shi2025_quadCascadeEKT}
identifies the resulting constant-$\Pi_\sigma$ subrange with a $-7/5$ exponent. Imposing this exponent fixes the saturation of $\sigma$, and hence of $f$,
\begin{equation}
\sigma_\infty^{\mathrm{EK}} = -\tfrac{4}{15},
\qquad
f_\infty = \sigma_\infty^{\mathrm{EK}} - a_3\,\mathrm{Re}^{-1/4}
        = -\tfrac{4}{15} - \tfrac{3}{20}
        = -\tfrac{5}{12},
\label{eq:f_Rae_sat}
\end{equation}
where the second equality uses $\mathrm{Re}=1$ for the Shi configuration. We
therefore adopt the minimal saturating form
\begin{equation}
f\!\left(Ra_e^{\,\mathrm{norm}}\right)
= -\frac{5}{12}
\left[
1 - \exp\!\left(-\left(\frac{Ra_e^{\,\mathrm{norm}}}{Ra_e^{\times}}\right)^{p}\right)
\right],
\label{eq:f_Rae}
\end{equation}

in which the crossover scale $Ra_e^{\times}$ and the dimensionless steepness $p=\mathcal{O}(1)$ are calibrated according to Shi~\textit{et al.} ($Ra_e^{\mathrm{norm}}=7.23\times10^{5}$,
slope $-5/3$; $Ra_e^{\mathrm{norm}}=1.13\times10^{6}$, slope $-7/5$, giving
$Ra_e^{\times}\approx8.5\times10^{5}$ and $p\approx5$). The bounded saturating form of $f(Ra_e^{\mathrm{norm}})$ remains compatible with the experimentally  observed high-$Ra_e$ steepening toward a $-7/5$ spectrum. The complete set of input parameters is summarized in Table~\ref{tab:summary}

Wang~\textit{et al.}~\cite{Wang2016_microEKT} also demonstrated turbulent-like flow 
at low Reynolds number by electrokinetically forcing two co-flowing streams 
separated by the same sharp conductivity gradient in a microdiffuser channel (5000:1) (AC electric field at 100kHz). 
Using LIFPA, they reported a 
mean streamwise velocity $U = 2.9~\mathrm{mm\,s^{-1}}$ at 
$V_\mathrm{pp} = 20~\mathrm{V}$, with turbulence intensity $37\%$, 
giving $u_\mathrm{rms} \approx 1.07~\mathrm{mm\,s^{-1}}$. With the reported definition of Taylor microscale (Eq~\ref{eq:Taylor microscale}), Taylor Reynolds number $Re_\lambda=0.03$ and $\nu=10^{-6}\mathrm{m^2\,s^{-1}}$, the Taylor microscale $\lambda\approx 28\mathrm{\mu m}$. 

To populate the master spectrum [Eq.~\eqref{eq:master}], the relevant 
parameters are determined as follows. The characteristic length is taken 
as the hydraulic diameter $D_h = 2wh/(w+h) \approx 168~\mathrm{\mu m}$, 
with $w = 130~\mathrm{\mu m}$ and $h = 240~\mathrm{\mu m}$. Applying the 
isotropic dissipation estimate (Eq.~\ref{eq:isotropic dissipation})
gives $\varepsilon \approx 2.2\times10^{-2}~\mathrm{m^2\,s^{-3}}$. 
The viscous cutoff wavenumber is estimated at $k_{\eta,spectrum}\approx 2\times10^{5}\mathrm{m^{-1}}$ from the spectral observation at $\eta\approx 5 \mathrm{\mu m}$. 
The bulk Reynolds number remains low throughout ($Re < 1 $), 
confirming that inertial contributions to the spectral slope are negligible 
and that the dominant non-inertial mechanism is electrokinetic forcing, 
characterized by electric Grashof number $(Gr_e) = 7.1\times10^5$. In the slope correction 
[Eq.~\eqref{eq:slope}], the Reynolds number term $a_3\,\mathrm{Re}^{-1/4}$ remains active and captures the low-Reynolds-number viscous character of the microfluidic flow, 
and $Gr_e$ is converted electric Rayleigh number through the relationship $Ra_e=ScGr_e$ ($Sc=v/D_e$) followed by normalization (See supporting information). 
The slope correction $\sigma$ gives a predicted slope = $-1.85$, which remains consistent with the $-5/3 \pm12\%$ power-law slope reported by Wang~\textit{et al.} 
over more than one decade in frequency~\cite{Wang2016_microEKT}.

The physics-specific cutoff ($k_\star$) is determined from the nominal electric 
Kolmogorov scale~\cite{Wang2016_microEKT}, $\ell_{ed} = w/\sqrt{Gr_e}$. 
For Shi~\textit{et al.}, electric Grashof number ($Gr_e$) is obtained from $Gr_e=Ra_e/Sc$, and the estimated $\ell_{ed}\approx 9.42\times10^{-6}~\mathrm{m}$ and $7.53\times10^{-6}~\mathrm{m}$ for $Ra_e=2.89\times10^{6}$ and $4.52\times10^{6}$. The corresponding physics-cutoff scale ($k_\star$) are $1.06\times10^{5}~\mathrm{m^{-1}}$ and $1.33\times10^{5}~\mathrm{m^{-1}}$. For Wang~\textit{et al.}, $\ell_{ed} \approx 1.54\times10^{-7}~\mathrm{m}$, giving $k_\star = 1/\ell_{ed} 
\approx 6.5\times10^6~\mathrm{m^{-1}}$. Notably, $k_\star \gg k_\eta$, 
indicating that viscous dissipation terminates the energy cascade well 
before the electrokinetic mechanism reaches its own characteristic cutoff 
scale. Consequently, the exponential viscous damping term governs the 
high-wavenumber rolloff in Eq.~\eqref{eq:master}, while $k_\star$ 
represents an upper bound that is not reached under the present 
experimental conditions.

The crossover wavenumber ($k_0$) is estimated by Eq.~\eqref{eq:k_0}. For electrokinetically driven microfluidic turbulence, we take $\ell_0=w_0$ as the characteristic forcing width,
\begin{equation}
k_0 \sim \frac{1}{w_0}.
\end{equation}
Thus, in the present electrokinetic cases, \(k_0\) coincides with the inverse channel-width scale, indicating that the cascade-onset length is set by the effective electric-field forcing distance. Using Eq.~\eqref{eq:k_0}, we obtain \(k_0=1.61\times10^{3}\,\mathrm{m^{-1}}\) for Shi \textit{et al.}, with \(w_0=620~\mu\mathrm{m}\), and \(k_0=7.69\times10^{3}\,\mathrm{m^{-1}}\) for Wang \textit{et al.}, with \(w_0=130~\mu\mathrm{m}\).

\begin{figure}[t]
    \centering
    
    \begin{subfigure}[b]{0.8\textwidth}
        \centering
        \includegraphics[width=\textwidth]{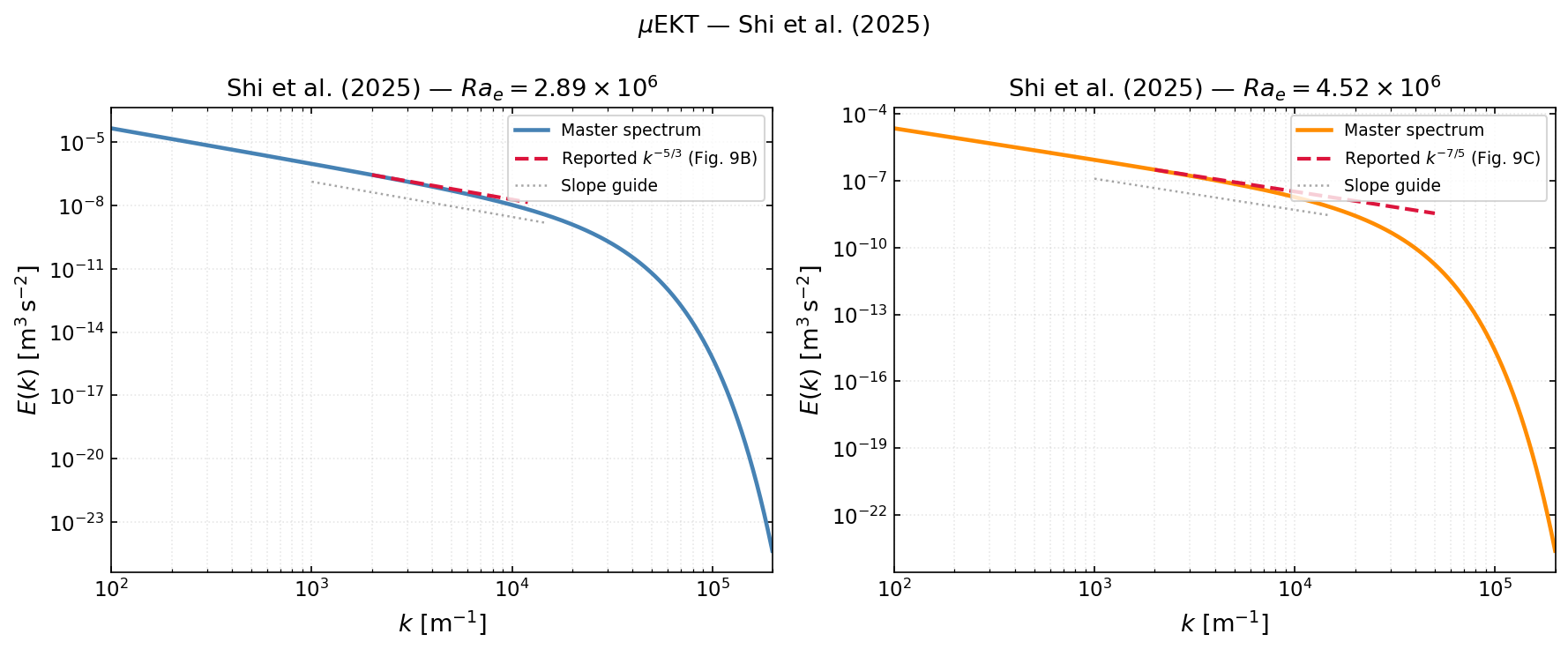}
        
        \label{fig:muEKT_shi}
    \end{subfigure}
    \hfill
    \begin{subfigure}[b]{0.48\textwidth}
        \centering
        \includegraphics[width=\textwidth]{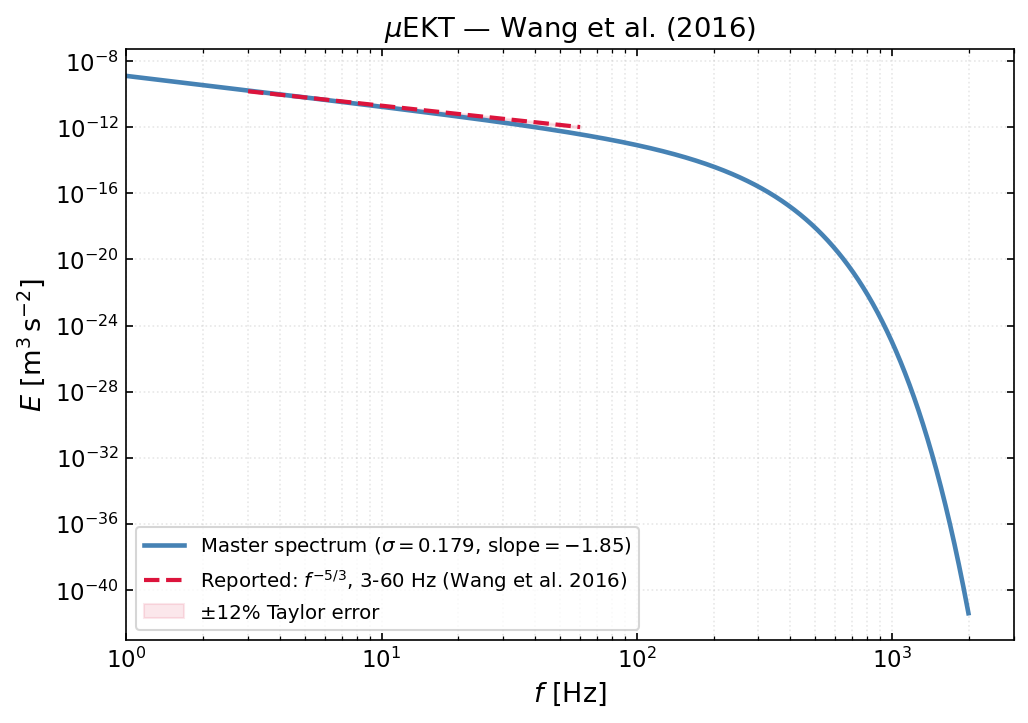}
        \label{fig:muEKT_wang}
    \end{subfigure}
    
    \caption{%
    Spectral scaling for microelectrokinetic turbulence ($\mu$EKT).
    (a)~Wavenumber spectrum from Shi~\textit{et al.}~\cite{Shi2025_quadCascadeEKT}:
    solid lines show the master spectrum at $Ra_e=2.89\times10^6$
    (slope~$-5/3$, $\sigma=-8.62\times10^{-4}$) and $Ra_e=4.52\times10^6$
    (slope~$-7/5$, $\sigma=-0.267$);
    dashed lines show the reported power-law scalings from Figs.~9B and~9C,
    normalised to the master spectrum at $k_0=1.61\times10^3$~m$^{-1}$.
    (b)~Frequency spectrum from Wang~\textit{et al.}~\cite{Wang2016_microEKT}
    at $V_f=20~\mathrm{V_{p-p}}$; the predicted slope $-1.85$ lies within the
    $\pm12\%$ Taylor-hypothesis uncertainty (shaded band) of the reported
    $f^{-5/3}$ scaling over 3--60~Hz.%
    }
    \label{fig:muEKT_comparison}
\end{figure}

The resulting master spectra for micro-electrokinetic turbulence, computed with $\gamma_v = 1$ and $\alpha_v = 4/3$, are shown in Fig.~\ref{fig:muEKT_comparison}. The predicted spectra capture both the electro-inertial-range slope and the onset of viscous rolloff at each $Ra_e$ condition without additional free parameters beyond those in Table~\ref{tab:summary}. Under low Reynolds number in microfluidic condition ($Re_\lambda \approx 1$), turbulence is motivated by non-inertial electrokinetic force instead of inertial force, and it necessitates a microfluidic modification to the classical Kolmogorov cascade assumptions; the present master spectrum and slope modification therefore serve as the phenomenological interpolation between the inertial and electrokinetically dominated limits, informed by the quad-cascade theory of Shi~\textit{et al.}

\subsubsection{Interface-induced turbulence}

Padhan~\textit{et al.}~\cite{Padhan2024_interfaceTurbulence} demonstrated 
a low-Reynolds-number, interface-induced turbulent state in viscous 
binary-fluid mixtures, in which interfacial fluctuations destabilize an 
otherwise laminar cellular flow. In their two-dimensional DNSs of the 
Cahn--Hilliard--Navier--Stokes (CHNS) equations, the flow is maintained 
at $Re=1$, well below the single-fluid instability threshold 
$Re_c=\sqrt{2}$~\cite{Gotoh1984_cellularInstability}, such that inertial 
instability is entirely excluded and the observed spatiotemporal chaos is 
sustained solely by interfacial stresses. The resulting nonequilibrium 
statistically steady state (NESS) exhibits a shell-averaged kinetic-energy 
spectrum with a power-law regime $E(k)\sim k^{-4.5}$.

To benchmark the master-spectrum framework against this regime, the 
nondimensional simulation parameters are mapped onto physical units by 
assigning the computational domain side $L_0=2\pi$ to a representative 
microfluidic length $L_{\mathrm{box}}=100~\mu\mathrm{m}$, characteristic 
of PDMS microchannel geometries used in experimental low-Reynolds-number 
turbulence studies~\cite{Wang2016_microEKT}. The cellular forcing at 
$k_f=4$ then corresponds to the physical injection wavenumber
\begin{equation}
k_0 = \frac{2\pi}{L_{\mathrm{box}}}\times k_f
\approx 2.51\times10^{5}~\mathrm{m}^{-1},
\label{eq:k0_interfacial}
\end{equation}
which sets the crossover wavenumber $k_0$ in Eq.~\eqref{eq:master}.

The viscous cutoff $k_\eta$ is estimated from the scale-by-scale energy 
budget in the inset of Fig.3(b) of Padhan~\textit{et al.}, where the 
viscous dissipation $2\nu k^2 E(k)$ overtakes the interfacial stress 
$S(k)$ at $k/k_f\approx5$--$8$. Taking the representative midpoint 
$k/k_f\approx7$ gives
\begin{equation}
k_\eta \approx 7\,k_0 \approx 1.76\times10^{6}~\mathrm{m}^{-1}.
\label{eq:keta_interfacial}
\end{equation}
The physics-specific cutoff $k_\star$ is determined from the inverse 
Cahn--Hilliard interfacial width $\xi$, the length scale that sets 
the spatial support of the interfacial stress 
$\mathbf{S}_\phi = -\phi\nabla(\delta F/\delta\phi)$~\cite{Padhan2024_interfaceTurbulence}. The stress is 
confined to the diffuse-interface layer of thickness $\xi$ over 
which the order parameter $\phi$ transitions between its two bulk 
values $\pm 1$, and vanishes in the bulk phases where $\phi$ is 
saturated and $\nabla\phi \to 0$. Because $\mathbf{S}_\phi$ is 
real-space localized over the width $\xi$, its spectral 
contribution $S(k)$ to the kinetic-energy budget decays for 
$k > 1/\xi$, beyond which the interfacial mechanism can no 
longer inject energy into the cascade.
The Cahn number 
$\mathrm{Cn}=\xi/L_{\mathrm{box}}=0.006$ is reported in the 
Supplemental Material of Padhan~\textit{et al.}~\cite{Padhan2024_interfaceTurbulence}, 
giving directly
\begin{equation}
\xi_{\mathrm{phys}} = \mathrm{Cn}\times L_{\mathrm{box}} 
= 0.006\times100~\mu\mathrm{m} 
= 6.00\times10^{-7}~\mathrm{m},
\end{equation}
and therefore
\begin{equation}
k_\star = \frac{1}{\xi_{\mathrm{phys}}}
\approx 1.67\times10^{6}~\mathrm{m}^{-1}.
\label{eq:kstar_interfacial}
\end{equation}
This value is consistent with the grid-resolution requirement stated 
in the main text --- that interfaces are resolved by approximately 
three grid points on the $512^2$ domain --- which independently gives 
$\xi_{\mathrm{nd}}\approx3\times(2\pi/512)\approx0.0368$, corresponding 
to $\mathrm{Cn}\approx0.00586$, within 2\% of the reported value. 
The resulting scale ratio $k_\star/k_0\approx6.6$ corresponds to 
less than one decade of cascade range, consistent with the narrow 
power-law region visible in Fig.3(b) of Padhan~\textit{et al.}


The dissipation rate $\varepsilon_{\mathrm{diss}}$ is not reported 
directly by Padhan~\textit{et al.} and is estimated from the 
forcing-scale energy balance. Using the paper's scaling 
$U=f_0/(\nu k_f^2)$ and $Re=UL/\nu$ with $L=k_f^{-1}$, the 
constraint $Re=1$ gives $f_0\sim\nu^2 k_f^3$ and $U\sim\nu k_f$, 
so that the reference injection rate yields
\begin{equation}
\varepsilon_{\mathrm{diss}} \sim \nu^3 k_f^4
\approx 4.0\times10^{3}~\mathrm{m^2\,s^{-3}},
\label{eq:eps_interfacial}
\end{equation}
adopting a water-like kinematic viscosity 
$\nu=10^{-6}~\mathrm{m^2\,s^{-1}}$. 
Because $Re=1$, no scale-separated inertial cascade exists, and the slope correction reduces to the interfacial-stress channel plus the residual microscale-viscous term, $\sigma = a_3Re^{-1/4}+ f(Ca)$, with all other channels inactive ($M\to0$, $\chi\to0$, $Wi^{norm}\to0$, $Ra_e\to0$).

The $Ca$-dependence of $\sigma$ is constructed from the phase 
behavior tabulated by Padhan~\textit{et al.}, who report spatiotemporally chaotic flow 
at $Ca=0.15$, $0.18$, $0.19$, and $0.20$, with the periodic states elsewhere across the tested range $Ca\in[0.01,0.60]$

\begin{figure}[t]
    \centering
    \includegraphics[width=\textwidth]{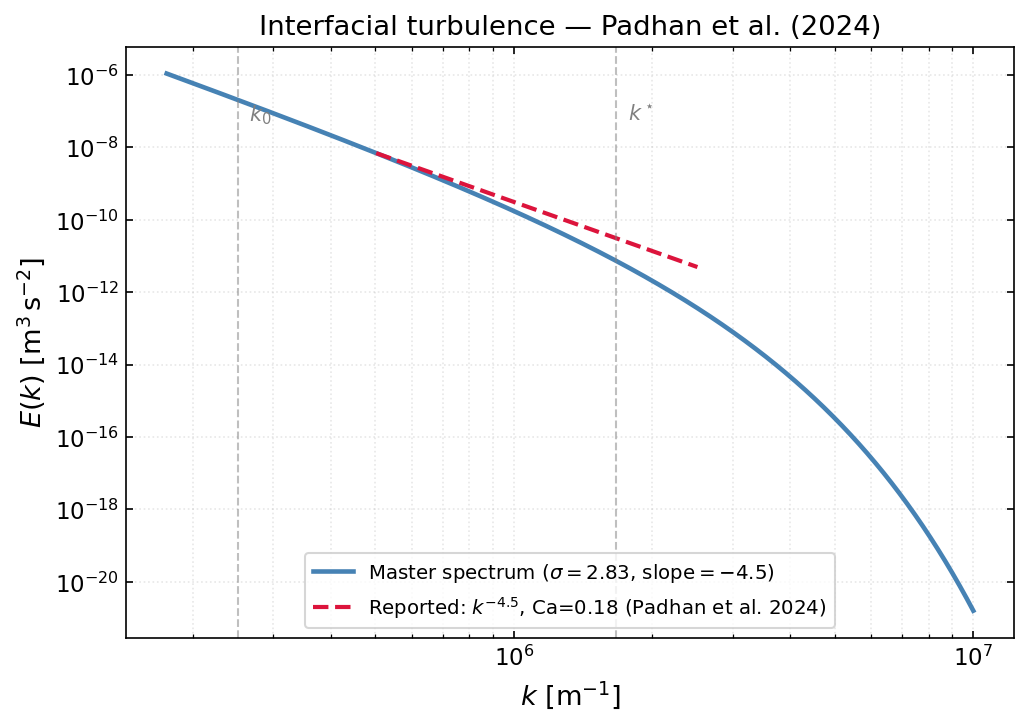}
    \caption{%
    Spectral scaling for interface-induced turbulence
    (Padhan~\textit{et al.}~\cite{Padhan2024_interfaceTurbulence}), $Ca=0.18$.
    Solid line: master spectrum ($\sigma=2.83$, slope $-4.5$).
    Dashed line: reported $k^{-4.5}$ scaling from Fig.~3b,
    normalised at $2k_0$ for slope comparison.
    Dotted vertical lines mark the injection wavenumber $k_0$ and
    physics-specific cutoff $k^\star$.%
    }
    \label{fig:interfacial_turbulence}
\end{figure}

The chaotic observations ($Ca=0.15$--$0.20$) define a narrow 
turbulent window, with periodic states bounding it below 
($Ca\leq0.12$) and above ($Ca\geq0.60$). We adopt a Gaussian bell form with a 
linear onset factor,
\begin{equation}
f(Ca) = A\; Ca \exp\!\left[
-\frac{(Ca - 0.18)^2}{2\omega^2}
\right],
\label{eq:f_Ca_gauss}
\end{equation}
which encodes $f(0)=0$ (no interface), a peak at $Ca=0.18$ 
(center of the chaotic window), and suppression outside the 
observed turbulent range. The width $\omega=0.025$ is chosen 
so that $f$ decays to below 10\% of its peak value by 
$Ca\approx0.22$, consistent with the last observed chaotic 
state at $Ca=0.20$ and the onset of periodic behavior beyond 
it. The prefactor $A=14.91$ is fixed by the single constraint 
$f(0.18)=161/60$, reproducing the reported spectral exponent 
$-5/3 - \sigma = -4.5$ at $Re=1$:
\begin{equation}
f(Ca) = 14.91\; Ca\exp\!\left[
-\frac{(Ca-0.18)^2}{2(0.025)^2}
\right].
\label{eq:f_Ca_final}
\end{equation}
The predicted exponent $-5/3-\sigma\approx-4.50$ at $Ca=0.18$ 
agrees with the reported $k^{-4.5}$ scaling, and 
Eq.~\eqref{eq:f_Ca_final} correctly suppresses the slope 
correction outside the observed turbulent window.
This construction is analogous to the $Ra_e$-dependent correction in 
the electrokinetic regime [Eq.~\eqref{eq:f_Rae}], where the functional 
form is likewise anchored to published spectral transitions. The 
predicted spectral exponent $-5/3-\sigma\approx-4.50$ agrees with the 
reported $k^{-4.5}$ scaling. 
The complete input parameters are summarized in 
Table~\ref{tab:summary}, and the resulting master spectra, computed with 
$\gamma_v=1$ and $\alpha_v=4/3$, are shown in Fig.~\ref{fig:interfacial_turbulence}.

\subsubsection{Active suspension turbulence}

Wensink \textit{et al.}~\cite{Wensink2012MesoScaleTurbulence} demonstrated 
that dense suspensions of motile \textit{B.~subtilis} generate self-sustained 
meso-scale turbulence despite operating at very low Reynolds number 
($Re \sim 10^{-4}$--$10^{-3}$), a regime in which inertial instabilities 
are entirely absent and the turbulent state is sustained solely by 
bacterial self-propulsion and steric interactions.
Using the \textit{B.~subtilis} body length $\ell \approx 5~\mu\mathrm{m}$ 
and a characteristic collective swimming speed 
$U \approx 100~\mu\mathrm{m\,s^{-1}}$~\cite{Wensink2012MesoScaleTurbulence}, 
the Reynolds number is
\begin{equation}
Re = \frac{U\ell}{\nu} \approx 5\times10^{-4},
\label{eq:Re_active}
\end{equation}
taking the kinematic viscosity of water $\nu = 10^{-6}~\mathrm{m^2\,s^{-1}}$.
At this Reynolds number, viscous forces dominate at every dynamically 
relevant scale simultaneously, so injection and dissipation are not 
scale-separated in the classical sense~\cite{Purcell1977_lowRe}.

Given the intrinsic directional forcing from the active micro-swimmer, flow anisotropy is assumed and 
the dissipation rate is estimated by
\begin{equation}
\varepsilon \sim \nu\frac{V^2}{\ell^2} 
\approx 4\times10^{-4}~\mathrm{m^2\,s^{-3}}.
\label{eq:eps_active}
\end{equation}
In active turbulence, kinetic energy is injected by bacterial 
self-propulsion at the swimmer body scale $\ell$, which sets the 
forcing wavenumber
\begin{equation}
k_0 = \frac{2\pi}{\ell} \approx 1.26\times10^{6}~\mathrm{m^{-1}}.
\label{eq:k0_active}
\end{equation}
The Wensink \textit{et al.} energy spectra show that 
turbulent vortices exist only for $k < k_\ell \equiv 2\pi/\ell$, 
and the spectrum rolls off sharply beyond this scale. Physically, direct active forcing is not expected to persist for
\(k > k_0\), because the stress generated by an individual bacterium
acts primarily over length scales comparable to its body size \(\ell\).
Accordingly, scales smaller than \(\ell\) are not taken to receive
sustained independent energy injection, and therefore the physics-specific dissipation scale coincides with the energy forcing scale:
\begin{equation}
k_{\star} = k_0 = \frac{2\pi}{\ell} \approx 1.26\times10^{6}~\mathrm{m^{-1}}.
\label{eq:kstar_active}
\end{equation}

Beyond the turbulent cascade, the cutoff is also contributed by viscous dissipation, which is determined by the smallest observable eddy scale directly from the reported energy spectrum given that $Re\ll1$~\cite{Wensink2012MesoScaleTurbulence}: 
\[
\frac{k_{\eta,spectrum}}{k_{\ell}} \approx 0.5,
\]
which gives
\[
k_{\eta,spectrum} \approx 6.28\times10^{5}~\mathrm{m^{-1}}.
\]

The slope correction $\sigma$ in Eq.~\eqref{eq:slope} 
receives contributions solely from the Reynolds correction 
term, with all other forcing parameters inactive 
($M \to 0$, $\chi \to 0$, $Wi^{norm} \to 0$, 
$Ra_e \to 0$, $Ca \to 0$). At $Re = 5\times10^{-4}$, 
the correction evaluates to
\begin{equation}
\sigma = a_3\,Re^{-1/4} 
= \frac{3}{20}\times(5\times10^{-4})^{-1/4} 
\approx 1.00,
\label{eq:sigma_active}
\end{equation}
giving a predicted spectral exponent
\begin{equation}
-\frac{5}{3} - \sigma \approx -\frac{8}{3},
\label{eq:slope_active}
\end{equation}
in agreement with the $k^{-8/3}$ large-$k$ 
scaling reported by Wensink \textit{et al.}. The result suggests that the 
extremely low $Re$ generated by bacterial 
self-propulsion at the microscale places the 
system deep in the viscosity-dominated regime, 
where the $a_3\,Re^{-1/4}$ correction grows 
large and accounts for the full deviation from 
the classical Kolmogorov $-5/3$ exponent

\begin{figure}[t]
    \centering
    \includegraphics[width=\textwidth]{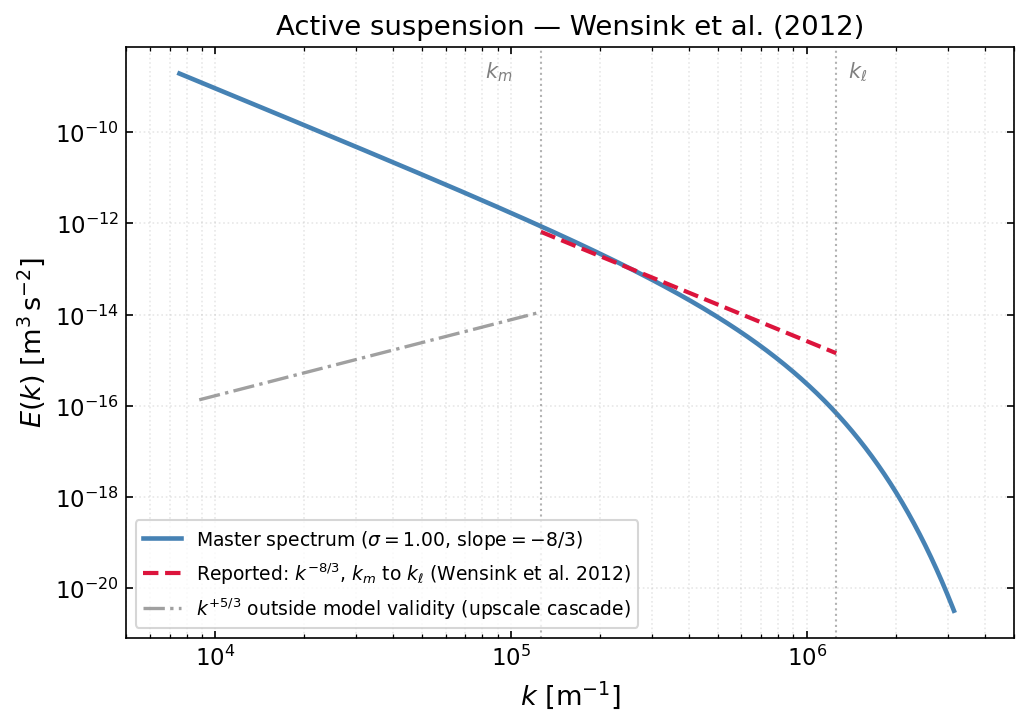}
    \caption{%
    Spectral scaling for active bacterial suspension
    (Wensink~\textit{et al.}~\cite{Wensink2012MesoScaleTurbulence}),
    $Re\approx5\times10^{-4}$.
    Solid line: master spectrum ($\sigma=1.00$, slope $-8/3$).
    Dashed line: reported $k^{-8/3}$ scaling from Fig.~4B,
    valid over the inverse-cascade window $k_m\leq k\leq k_\ell$.
    Dash-dot grey line: $k^{+5/3}$ at $k<k_m$, arising from upscale
    energy transfer; this lies outside the master spectrum
    validity window.
    Dotted vertical lines mark $k_m$ and $k_\ell=2\pi/\ell$.%
    }
    \label{fig:Active_turbulence}
\end{figure}

The damping parameters are set to $\gamma_v = 1$, 
$\alpha_v = 4/3$ for the viscous term and $\gamma_p = 1$, 
$\alpha_p = 1$ for the physics-specific term, consistent 
with the values used in the electrokinetic and interfacial 
regimes. The complete input parameters are summarized in 
Table~\ref{tab:summary}, and the resulting predicted 
spectrum is shown in Fig.~\ref{fig:Active_turbulence}.
The $k^{+5/3}$ behaviour observed at $k < k_m$ in Fig.~4B of
Wensink~\textit{et al.}~\cite{Wensink2012MesoScaleTurbulence},
arising from upscale energy transfer toward the mesoscale vortex
at $R_v\approx10\ell$, lies outside the master spectrum's validity
window $k_m\leq k\leq k_\ell$ [Eq.~\eqref{eq:bidirectional_topology}]
and is not described by the present framework.

\subsubsection{Elastic turbulence}

The seminal experiment of Groisman and Steinberg~\cite{Groisman2000_elasticTurbulence}
established that a dilute solution of high-molecular-weight polyacrylamide
driven in a swirling flow between two parallel disks develops a chaotic,
multi-scale velocity field at small Reynolds number ($Re<1$).
The associated velocity power spectrum decays as a power law,
$E(k)\sim k^{-\xi}$, with exponent $\xi>3$---markedly steeper than the
classical Kolmogorov scaling $k^{-5/3}$ and below the dimensional bound
$k^{-3}$ characteristic of inertia-dominated cascades. This spectral
signature reflects the inertialess, polymer-stress-driven nature of
elastic turbulence (ET): the kinetic-energy cascade is replaced by a
flux of elastic energy stored in the stretched polymer chains and
dissipated through polymer relaxation~\cite{FouxonLebedev2003}.

Building on the Oldroyd-B constitutive description of dilute polymer
solutions, we model the spectral exponent as a baseline Kolmogorov
slope corrected by an elastic modifier $\sigma$, whose magnitude is
set by the local balance between polymer stretching and relaxation.
This balance is encapsulated by the Weissenberg number
$\mathrm{Wi}=\lambda\dot{\gamma}$, where $\lambda$ is the polymer
relaxation time and $\dot{\gamma}$ a characteristic shear rate.
The elastic contribution to the slope is written
\begin{equation}
f(\mathrm{Wi}^{\mathrm{norm}})
= \sigma_{\mathrm{peak}}^{\mathrm{ET}}
\Bigl[\,1-\exp\!\bigl(-\kappa\,\mathrm{Wi}^{\mathrm{norm}}\bigr)\Bigr],
\label{eq:f_Wi_general}
\end{equation}
with saturation value $\sigma_{\mathrm{peak}}^{\mathrm{ET}}\approx 11/6$
and steepness parameter $\kappa\gg 1$ calibrated from the swirling-flow experiment of Groisman and Steinberg~\cite{Groisman2000_elasticTurbulence}, which reports a velocity power
spectrum of slope $-3.5$ in the developed regime ($\mathrm{Wi}=13$,
$\mathrm{Wi}_c=3.5$, $\mathrm{Wi}^{\mathrm{norm}}\approx2.7$, in the saturated
branch). The normalized Weissenberg number
\begin{equation}
\mathrm{Wi}^{\mathrm{norm}}
= \frac{\mathrm{Wi}-\mathrm{Wi}_{c}}{\mathrm{Wi}_{c}}
\label{eq:Wi_norm}
\end{equation}
removes the geometry-specific onset threshold $\mathrm{Wi}_{c}$, so that
data from different channel geometries can be compared on a single
master curve. Equation~\eqref{eq:f_Wi_general} embeds two physical
requirements: (i) below onset ($\mathrm{Wi}^{\mathrm{norm}}\!\leq\!0$)
the modifier vanishes and the spectrum reverts to its Newtonian
baseline; and (ii) once the polymer chains are fully extended and the
stretching/relaxation balance reaches its asymptote, $\sigma$ saturates
at $\sigma_{\mathrm{peak}}^{\mathrm{ET}}=11/6$, yielding a total slope
\begin{equation}
\xi_{\mathrm{ET}}
= \tfrac{5}{3}+\tfrac{11}{6}
= \tfrac{7}{2},
\label{eq:slope_total}
\end{equation}
i.e.\ a velocity spectrum $E(k)\sim k^{-7/2}$.

In the elastic regime the inertial contribution to the slope modifier,
which scales as $a_{3}\,\mathrm{Re}^{-1/4}$ in classical drag-reduction
frameworks, vanishes ($a_{3}\,\mathrm{Re}^{-1/4}\!\rightarrow\!0$), so
the slope is controlled entirely by elasticity,
$\sigma_{\mathrm{ET}}=f(\mathrm{Wi}^{\mathrm{norm}})$. This limit is
consistent with the inertialess, polymer-stress-driven cascade of
Fouxon and Lebedev~\cite{FouxonLebedev2003}: the cascade is not an
inertial range modified by elasticity, but an inertialess chaotic state
in which the polymer elastic stress and its relaxation set the spectral
structure.

To populate and test Eq.~\eqref{eq:f_Wi_general} across distinct flow
geometries, we draw on two independent experimental studies that
report well-resolved ET spectra: a planar cross-slot microchannel
\cite{Sousa2018_crossslot} and a cubically packed porous bed
\cite{Carlson2022_porousET}.

\emph{Predictions (packed porous bed).} Carlson \textit{et al.}~\cite{Carlson2022_porousET} confined a cubically
packed array of glass spheres (lattice pitch $a\!=\!\SI{500}{\micro\meter}$)
inside a microchannel and circulated a high-molecular-weight
polyacrylamide solution in a water/glycerol solvent
($\sim\!89.5~\mathrm{wt\%}$ glycerol). An apparent flow instability is
observed at $\mathrm{Wi}_{c}\!=\!1.8$, and a sustained ET state is
reached at $\mathrm{Wi}\!=\!8.2$
($\mathrm{Wi}^{\mathrm{norm}}\!\approx\!3.6$, again in the saturated
branch). The lattice pitch ($w_p$) sets the energy-injection scale,
$k_{0}=1/w_p \approx\!\SI{2e3}{\per\meter}$. The dissipation rate is
estimated as
\begin{equation}
\varepsilon
\approx \frac{\sigma_{\mathrm{tot}}\dot{\gamma}}{\rho}
\approx \SI{1.99e-2}{\meter\squared\per\second\cubed},
\label{eq:eps_Carlson}
\end{equation}
following Bird \textit{et al.}~\cite{Bird1987_polymerLiquids}.

Because $\mathrm{Re}\!\sim\!\mathcal{O}(10^{-3})$ places the Kolmogorov
viscous cutoff far above the experimentally accessible range, we
identify the high-wavenumber cutoff directly from the spectral
roll-off. Introducing the dimensionless frequency $f^{\star}=f\lambda$
and invoking Taylor's frozen-turbulence hypothesis with pore-scale
advection velocity $U_{p}$ yields
\begin{equation}
k_{\eta,spectrum}
= \frac{2\pi\,(f^{\star}/\lambda)}{U_{p}}
\approx \SI{2.45e5}{\per\meter},
\qquad f^{\star}\!\approx\!80.
\label{eq:k_eta_PSD}
\end{equation}

The physical (elastic) cutoff $k_{\star}$, which marks the upper bound of 
the polymer-stress-dominated band, is set by the polymer-physics floor 
on the elastic cascade: below the fully extended contour length 
$R_{\max}$ of a single polymer chain, no individual molecule can 
sustain coherent elastic stress, and the continuum viscoelastic 
description (Oldroyd-B, FENE-P) ceases to apply~\cite{Bird1987_polymerLiquids,
larson1999structure}. For a vinyl polymer in the 
all-trans (zigzag) backbone configuration, the contour length is
\begin{equation}
    R_{\max} = \frac{M_w}{M_0}\,b,
    \label{eq:Rmax}
\end{equation}
where $M_w$ is the polymer molecular weight, $M_0$ is the monomer 
molecular weight, and $b = 2\,d_{\mathrm{CC}}\sin(\theta/2) 
\approx 0.252~\mathrm{nm}$ is the per-monomer projection of two 
backbone C--C bonds (bond length $d_{\mathrm{CC}} = 0.154~\mathrm{nm}$, 
tetrahedral bond angle $\theta = 109.5^{\circ}$) onto the chain 
axis~\cite{rubinstein_colby_2003,flory1969statistical}. Equation 
\eqref{eq:Rmax} defines a fluid-specific elastic cutoff that is 
independent of flow geometry or Weissenberg number.

Both the Carlson~\cite{Carlson2022_porousET} and 
Sousa~\cite{Sousa2018_crossslot} experiments use the same viscoelastic 
fluid (polyacrylamide, $M_w = 18 \times 10^{6}~\mathrm{g\,mol^{-1}}$, 
Polysciences) with $M_0 \approx 71~\mathrm{g\,mol^{-1}}$ and 
$b \approx 0.25~\mathrm{nm}$, giving $R_{\max} \approx 64~\mu\mathrm{m}$ 
and
\begin{equation}
    k_{\star} = \frac{1}{R_{\max}} \approx 1.56 \times 10^{4}~\mathrm{m^{-1}}.
    \label{eq:kstar_Rmax}
\end{equation}
This sets the small-scale termination of the elastic cascade in both 
datasets and is consistent with the observation that ET spectra exhibit 
a clean power-law range only up to a sub-geometric scale of comparable 
magnitude.

\emph{Predictions (cross-slot).} Sousa \textit{et al.}~\cite{Sousa2018_crossslot} employed a planar
cross-slot device with aspect ratio (width:height) $1:1.1$, seeded with
high-molecular-weight polyacrylamide dissolved in a water--glycerol
solvent. Time-dependent (unsteady) flow first emerges near
$\mathrm{Wi}_{c}\!\approx\!6$, and the velocity spectrum steepens to a
slope of $-3.5$ for $\mathrm{Wi}\gtrsim 25.2$, corresponding to
$\mathrm{Wi}^{\mathrm{norm}}\!\approx\!3.2$---well inside the saturated
branch of Eq.~\eqref{eq:f_Wi_general}. The channel width
$L\!=\!\SI{100}{\micro\meter}$ sets the energy-injection scale, giving
$k_{0}=1/L\!\approx\!\SI{1e4}{\per\meter}$. With
$\varepsilon\!\approx\!\sigma_{\mathrm{tot}}\dot{\gamma}/\rho
\!\approx\!\SI{1.79e-1}{\meter\squared\per\second\cubed}$
\cite{Bird1987_polymerLiquids}, the viscous cutoff and elastic crossover
are $k_{\eta,spectrum}\!\approx\!\SI{1.5e4}{\per\meter}$ and
$k_{\star}\!\approx\!\SI{1.56e4}{\per\meter}$, respectively. The model
prediction $\xi_{\mathrm{ET}}=-7/2$ from Eq.~\eqref{eq:slope_total}
agrees with the experimentally measured exponent. The parameters are summarized in Table ~\ref{tab:summary}

\begin{figure}[tb]
\centering
 \includegraphics[width=0.45\linewidth]{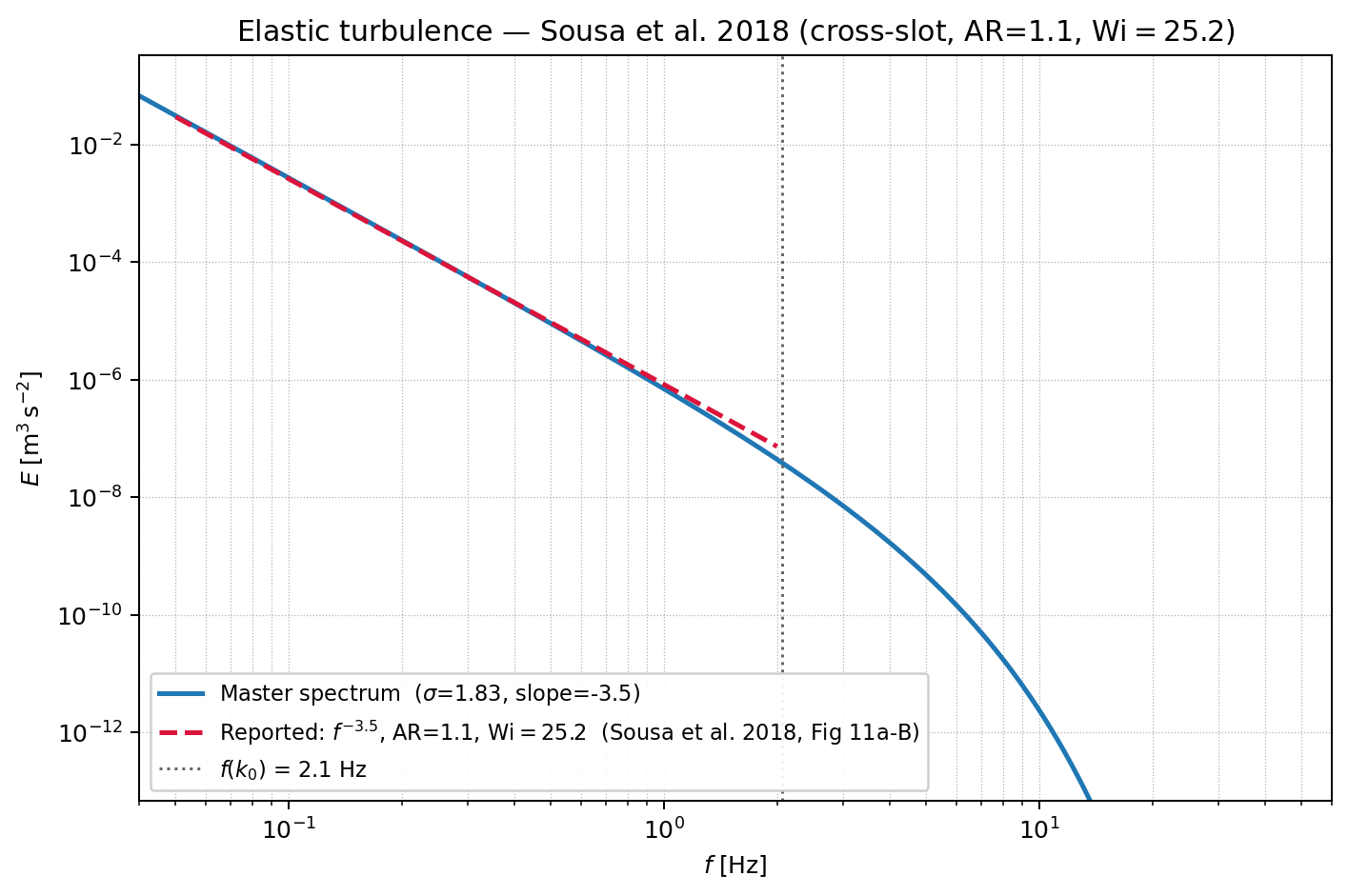}
  \includegraphics[width=0.45\linewidth]{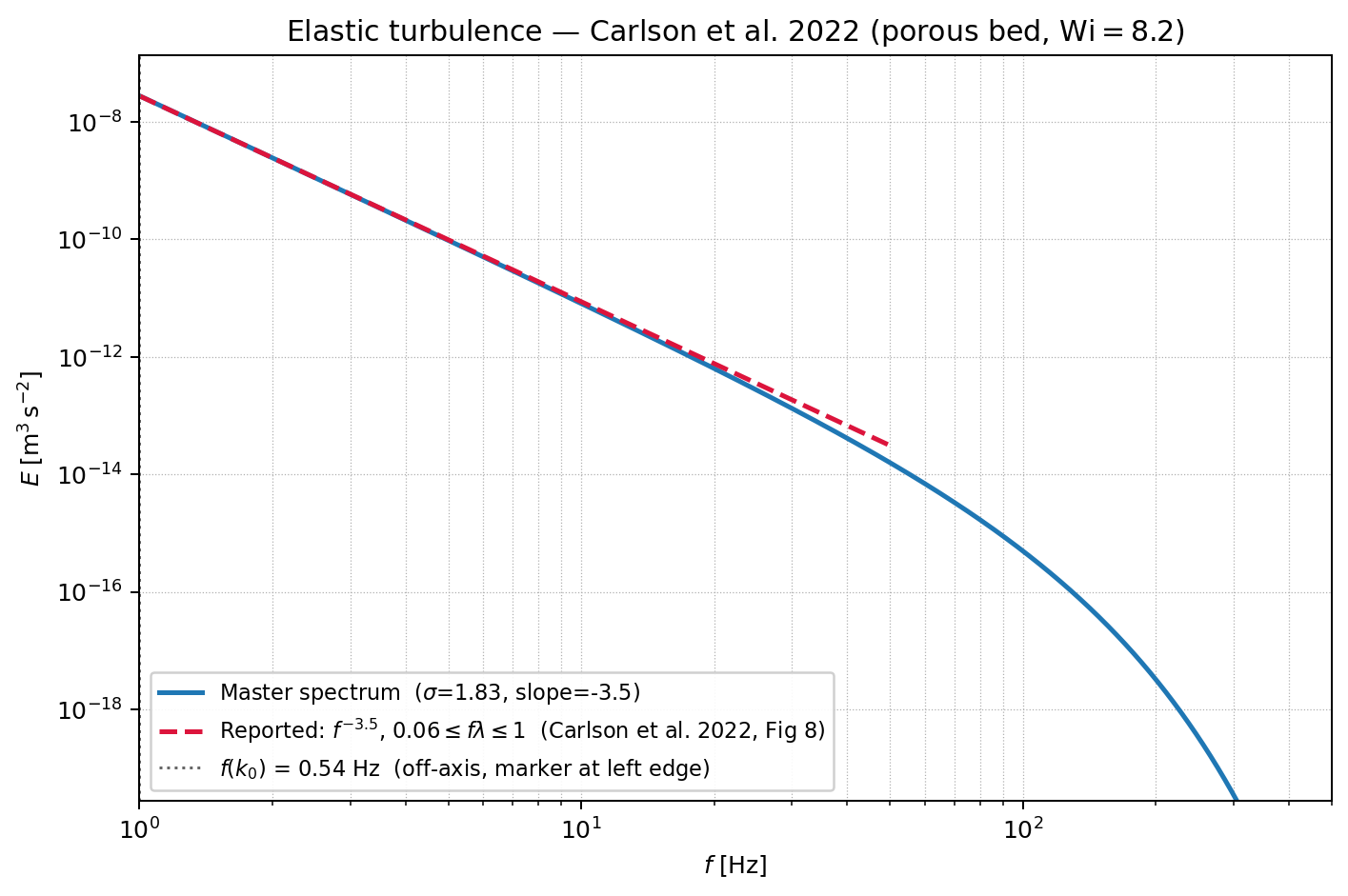}
\caption{Spectral scaling for elastic turbulence. (a)~ Spectrum from Sousa~\textit{et al.}~\cite{Sousa2018_crossslot}:
    solid lines show the master spectrum at $Wi=25.2$
    (slope~$-3.5$, $\sigma=1.83$);
    dashed lines show the reported power-law scalings from Fig.~8,
    normalised to the master spectrum at $k_0$.
    (b)~Spectrum from Carlson~\textit{et al.}~\cite{Carlson2022_porousET}
    at $Wi=8.2$; the predicted slope $-3.5$, $\sigma=1.83$;
    dashed lines show the reported power-law scalings from Fig.~11(a) (B curve), normalised to the master spectrum at $k_0$.}

\label{fig:master_curve}
\end{figure}

\begin{table*}
\caption{Control parameters for the five benchmark cases: $\mu$EK turbulence
from Shi~\textit{et al.}~\cite{Shi2025_quadCascadeEKT}, $\mu$EK turbulence from
Wang~\textit{et al.}~\cite{Wang2016_microEKT}, interfacial turbulence from
Padhan~\textit{et al.}~\cite{Padhan2024_interfaceTurbulence}, and active-suspension turbulence
from Wensink~\textit{et al.}~\cite{Wensink2012MesoScaleTurbulence}.}
\label{tab:summary}
\begin{ruledtabular}
\begin{tabular}{l c c c c c c c c c}
Case & $\varepsilon$ & $k_\eta$ & $k_0$ & $C_K$ & $\mathrm{Re}$ & $M$ &
$\mathrm{Wi}_\mathrm{norm}$ & $\mathrm{Ca}$ & $\mathrm{Ra}_{e,\mathrm{norm}}$ \\
\hline
$\mu$EKT (Shi), $\mathrm{Ra}_e=2.89\times10^{6}$
 & $2.98\times10^{-2}$ & $1.31\times10^{4}$ & $1610$ & $1$ & $1$ & $0$ & $0$ & $0$ & $7.23\times10^{5}$ \\
$\mu$EKT (Shi), $\mathrm{Ra}_e=4.52\times10^{6}$
 & $3.28\times10^{-2}$ & $1.35\times10^{4}$ & $1610$ & $1$ & $1$ & $0$ & $0$ & $0$ & $1.13\times10^{6}$ \\
$\mu$EKT (Wang), $20\,\mathrm{V_{pp}}$
 & $2.2\times10^{-2}$ & $2\times10^{5}$ & $7690$ & $1$ & $0.5$ & $0$ & $0$ & $0$ & $1.90\times10^{5}$ \\
Interfacial (Padhan), $\mathrm{Ca}=0.18$
 & $4.0\times10^{3}$ & $1.76\times10^{6}$ & $2.50\times10^{5}$ & $1$ & $1$ & $0$ & $0$ & $0.18$ & $0$ \\
Active suspension (Wensink)
 & $4.0\times10^{-4}$ & $6.28\times10^{5}$ & $1.26\times10^{6}$ & $1$ & $5\times10^{-4}$ & $0$ & $0$ & $0$ & $0$ \\
Elastic (Carlson porous), $Wi=8.2$,
 & $1.99\times10^{-2}$ & $2.45\times10^{5}$ & $2\times10^{3}$ & $1$ & $10^{-3}$ & $0$ & $3.6$ & $0$ & $0$ \\
 Elastic (Sousa cross-slot), $Wi=25.2$,  & $1.79\times10^{-1}$ & $1.5\times10^{4}$ & $1\times10^{4}$ & $1$ & $4.9\times10^{-4}$ & $0$ & $3.2$ & $0$ & $0$ \\
\end{tabular}
\end{ruledtabular}
\end{table*}

\begin{table}
\caption{Fitted spectral exponents, crossover wavenumber $k^\star$, and
integrated spectral-budget residual $\sigma$ as the spectral fingerprint for the cases of
Table~\ref{tab:summary}.}
\label{tab:exponents}
\begin{ruledtabular}
\begin{tabular}{l c c c c c c}
Case & $\gamma_v$ & $\alpha_v$ & $\gamma_p$ & $\alpha_p$ & $k^\star$ & $\sigma$ \\
\hline
$\mu$EKT (Shi), $\mathrm{Ra}_e=2.89\times10^{6}$
 & $1$ & $4/3$ & $1$ & $1$ & $1.06\times10^{5}$ & $-8.62\times10^{-4}$ \\
$\mu$EKT (Shi), $\mathrm{Ra}_e=4.52\times10^{6}$
 & $1$ & $4/3$ & $1$ & $1$ & $1.33\times10^{5}$ & $-0.267$ \\
$\mu$EKT (Wang), $20\,\mathrm{V_{pp}}$
 & $1$ & $4/3$ & $0$ & $1$ & $6.48\times10^{6}$ & $0.179$ \\
Interfacial (Padhan), $\mathrm{Ca}=0.18$
 & $1$ & $4/3$ & $1$ & $1$ & $1.67\times10^{6}$ & $2.83$ \\
Active suspension (Wensink)
 & $1$ & $4/3$ & $1$ & $1$ & $1.26\times10^{6}$ & $1.00$ \\
Elastic turbulence (Carlson porous)
 & $1$ & $4/3$ & $1$ & $1$ & $1.56\times10^{4}$ & $1.83$ \\
 Elastic turbulence (Sousa cross-slot)
 & $1$ & $4/3$ & $1$ & $1$ & $1.56\times10^{4}$ & $1.83$ \\
\end{tabular}
\end{ruledtabular}
\end{table}

\begin{table*}[tbp]
\caption{%
Quantitative comparison of predicted and reported spectral slopes for the five
benchmark cases.
The predicted slope $-5/3-\sigma$ uses inputs from Tables~\ref{tab:summary}
and~\ref{tab:exponents}.
The reported slope is taken directly from the cited work.
Percentage error: $100\times|m_\mathrm{pred}-m_\mathrm{rep}|/|m_\mathrm{rep}|$.
For Wang et al.\ and Carlson et al. \& Sousa et al., the comparison is in frequency space
(Taylor hypothesis applied); Wang et al.\ carries $\pm12\%$ Taylor uncertainty ~\cite{Wang2016_microEKT}.
}
\label{tab:slope_comparison}
\begin{ruledtabular}
\begin{tabular}{l c c c c}
Case & $\sigma$ & Predicted slope & Reported slope & Error (\%) \\
\hline
$\mu$EKT (Shi), $Ra_e=2.89\times10^6$ & $-8.62\times10^{-4}$ & $-1.67$ & $-5/3\approx-1.67$ & $<0.1$ \\
$\mu$EKT (Shi), $Ra_e=4.52\times10^6$ & $-0.267$             & $-1.40$ & $-7/5=-1.40$       & $<0.1$ \\
$\mu$EKT (Wang), $V_f=20~\mathrm{V_{p\text{-}p}}$ & $0.179$ & $-1.85$ & $-5/3\approx-1.67$ & $11\%~(\leq12\%\text{ Taylor})$ \\
Interfacial (Padhan), $Ca=0.18$        & $2.83$               & $-4.50$ & $-4.5$             & $<0.1$ \\
Active suspension (Wensink), $Re\approx5\times10^{-4}$ & $1.00$ & $-2.67$ & $-8/3\approx-2.67$ & $<0.1$ \\
Elastic turbulence (Carlson), $Wi=8.2$ & $1.83$               & $-3.50$ & $-3.5$             & $<0.1$ \\
Elastic turbulence (Sousa), $Wi=25.2$ & $1.83$               & $-3.50$ & $-3.5$             & $<0.1$ \\
\end{tabular}
\end{ruledtabular}
\end{table*}

\section{Discussion}
\subsection{Prediction spectrum interpretation}
The master spectrum, Eq.~\eqref{eq:master}, provides a compact and
unifying framework for interpreting turbulence across non-inertial
microfluidic regimes and for guiding experimental design.
The spectral slope encodes the regime-specific forcing physics: Classical Kolmogorov turbulence gives the canonical scaling
$E(k)\sim k^{-5/3}$ in the inertial constant-flux range, where viscous effects
are negligible during interscale energy transfer. By contrast, non-inertially
driven microfluidic turbulence develops at low Reynolds number, where viscous
damping remains active across turbulence cascade scales. This persistent
viscous friction provides an additional sink of kinetic fluctuations, causing the
spectrum to decay more rapidly toward smaller scales~\cite{Verma2022_variableEnergyFlux}. The resulting steepening
relative to the Kolmogorov slope is described by a regime-dependent exponent
$\sigma$, reflecting the balance between the specific non-inertial driving
mechanism and viscous dissipation.
The slope $-5/3 - \sigma$ thus constitutes a spectral fingerprint ($\sigma$ in Table~\ref{tab:exponents}) permitting regime identification directly from measured spectra without
recourse to flow visualisation or force decomposition.
The spectral slope modifier \(\sigma\) in Eq.~\eqref{eq:master} is expressed in terms of regime-specific dimensionless groups because such groups represent, by construction, the ratio of the non-inertial driving force to the viscous restoring force --- the same conceptual role that the Reynolds number plays in classical inertial turbulence.

In classical turbulence, the Reynolds number \(\mathrm{Re}\) quantifies the relative importance of inertial and viscous forces: at low \(\mathrm{Re}\), viscous damping suppresses perturbations and the flow remains laminar; above a critical value, inertial advection amplifies perturbations into eddies and the flow transitions to a turbulent state~\cite{Reynolds1883}. When \(\mathrm{Re}\) is sufficiently large, scale separation between forcing and dissipation gives rise to an inertial subrange in which the energy spectrum follows the universal Kolmogorov scaling $E(k)\sim k^{-5/3}$ ~\cite{Kolmogorov1941_localStructure}. The success of this single-parameter description rests on the fact that, in classical turbulence, all relevant physics reduces to the competition between inertial transfer and viscous dissipation, and \(\mathrm{Re}\) is the unique dimensionless group that captures this balance.

When a non-inertial mechanism replaces or supplements inertia as the driver, the energy budget acquires additional flux pathways [Eq.~\eqref{eq:flux_decomposition}], and a single dimensionless group is no longer sufficient. Each non-inertial mechanism introduces its own characteristic timescale and forcing rate, and the corresponding dimensionless groups --- the Weissenberg number for elastic stresses, the electric Rayleigh number for electrokinetic forcing, the Capillary number for interfacial stresses --- quantify the strength of that mechanism relative to viscous damping. 

Treating the cumulative effect of these non-inertial mechanisms as an additive shift to the spectral slope
($-5/3 \to -5/3-\sigma$) is supported by published results across several distinct regimes. In polymer-laden turbulence, the kinetic energy spectrum obeys \(E(k)\sim k^{-\alpha}\) with \(4<\alpha<5\), and the exponent \(\alpha\) decreases continuously with increasing Weissenberg number~\cite{Watanabe2014_powerLawSpectraPolymers} --- a direct numerical demonstration that the spectral slope varies systematically with the controlling dimensionless group, deviating from the Kolmogorov \(-5/3\) value in non-inertial driving regimes.

The overall decay prescribed by the spectral slope is bounded at small scales by two cutoff wavenumbers, $k_{\eta}$ and $k_{\star}$, which encode distinct competitions between viscous dissipation and the dominant nonlinear transfer mechanism in the flow. The relevant mechanism, and therefore the relevant cutoff, depends on the operating Reynolds number.

The viscous cutoff $k_{\eta}=\eta^{-1}$ marks the wavenumber at which viscosity arrests the downscale transfer of fluctuation energy sustained by the inertial nonlinearity $(\mathbf{u}\cdot\nabla)\mathbf{u}$. This nonlinearity is intrinsic to the Navier--Stokes equation and arises from the velocity field's self-transport of momentum, independent of how the flow is initially excited. Above $k_{\eta}$, the viscous damping rate $\nu k^{2}$ exceeds the inertial transfer rate $u(\ell)/\ell$, and the cascade flux is exhausted. The crossover is set by

\begin{equation}
\nu k^{2} \sim u(\ell)\,k,  k=\frac{1}{\ell}
\label{eq:inertial_cutoff}
\end{equation}

which, combined with the constant-flux condition $\varepsilon$, yields the Kolmogorov microscale (Eq.~\ref{eq:Kolmogorov scale}). Although introduced by Kolmogorov~\cite{Kolmogorov1941_localStructure,Pao1965_spectrum} in the context of the inertial cascade,

\begin{equation}
\eta = \left(\frac{\nu^{3}}{\varepsilon}\right)^{1/4},
\end{equation}

follows from dimensional analysis alone at $Re=1$---it is the unique length constructible from $\nu$ and $\varepsilon$---and therefore retains meaning whenever an inertial flux is well defined. The competition encoded in $k_{\eta}$ is between viscosity and the inertial advection flux arriving from larger scales; $k_{\eta}$ is defined downstream of, and contingent upon, the existence of that flux.

When $\mathrm{Re}(\ell)\gg 1$ across the relevant scales, the inertial flux is well established, $k_{\eta}$ is given by the Kolmogorov microscale, and the spectral rolloff occurs at the same wavenumber. When $\mathrm{Re}(\ell)\lesssim 1$ across the spectrum, however, the inertial subrange collapses: the viscous rate exceeds the inertial turnover rate at every scale,

\begin{equation}
u(\ell)k \ll \nu k^{2},
\end{equation}

and inertial self-advection cannot sustain a downscale cascade. In this regime the Kolmogorov identity ($\eta$) is no longer guaranteed to coincide with the observed spectral rolloff, and $k_{\eta}$ is taken directly from the high-wavenumber termination of the measured velocity spectrum. This operational definition reduces to the Kolmogorov form in the high-Reynolds-number limit and is, in general, regime-specific at $\mathrm{Re}\ll 1$.

The physics cutoff $k_{\star}$, defined by

\begin{equation}
\tau_{\mathrm{regime}}(k_{\star})
=
\tau_{\mathrm{viscous}}(k_{\star}),
\label{eq:k_star}
\end{equation}

encodes a categorically different competition: viscosity against nonlinear non-inertial advection. The relevant nonlinearity is not the quadratic self-interaction of the velocity field but the two-way coupling between $\mathbf{u}$ and an external scalar field---the conductivity gradient sustaining electrokinetic body forces, the polymer conformation field driving elastic stresses, the swimmer orientation field sustaining dipolar flows, or the phase field setting interfacial stress. In each case, $\mathbf{u}$ advects the scalar while the scalar simultaneously forces $\mathbf{u}$, a coupled quadratic nonlinearity that is non-inertial in origin and that injects fluctuations across a range of scales rather than at the large scale alone. Above $k_{\star}$, viscous smoothing is fast enough to suppress these fluctuations at the moment of injection, before any cascade can develop.

The two cutoffs therefore demarcate distinct sources of energy input to the spectral budget: $k_{\eta}$ governs where viscosity exhausts the inertial self-advection flux propagating downscale from the forcing range, while $k_{\star}$ governs where viscosity suppresses the non-inertial scalar-driven injection at its source. Both have the structural form

\begin{equation}
\nu k^{2}
\sim
\tau^{-1}(k),
\label{eq:cutoff_generic}
\end{equation}

but the competing rates $\tau^{-1}$ are categorically different: for $k_{\eta}$, $\tau^{-1}$ is the inertial turnover rate, an intrinsic property of the velocity field; for $k_{\star}$, $\tau^{-1}$ is the non-inertial forcing rate, a property of the driving scalar and its coupling to the flow. In regimes where $\mathrm{Re}(\ell)\ll 1$ and the non-inertial mechanism dominates injection across all scales, the two cutoffs need not be separated and may coincide; in regimes where an inertial subrange persists alongside non-inertial forcing, they are independent parameters in the master spectrum.

The wavenumber \(k_0\) entering Eq.~\eqref{eq:master} plays a role distinct from the cutoff scales \(k_\eta\) and \(k_\star\). Whereas \(k_\eta\) and \(k_\star\) mark the wavenumbers at which viscous dissipation overtakes, respectively, the inertial self-advection flux and the non-inertial scalar-driven injection, \(k_0\) identifies the scale at which the regime-specific forcing mechanism deposits energy into the velocity field. We define $k_0$ regime-specifically through the geometric or material scale of the forcing: the inverse conductivity gradient width ($1/w_0$) for electrokinetic turbulence ~\cite{Wang2016_microEKT}, the inverse swimmer body length for active suspensions ($2\pi/\ell$)~\cite{Wensink2012MesoScaleTurbulence}, the externally imposed cellular forcing wavenumber for interfacial flows ($2\pi k_f/L_{box}$)~\cite{Padhan2024_interfaceTurbulence}

Two qualifications on this construction are necessary. First, \(k_0\) does not in general bound the spectrum from below. In forward-cascade regimes [Eq.~\eqref{eq:forward_topology}] the cascade proceeds monotonically from \(k_0\) toward \(\min(k_\eta,k_\star)\), and \(k_0\) coincides with the spectral lower edge. In bidirectional regimes [Eq.~\eqref{eq:bidirectional_topology}], by contrast, energy injected at \(k_0\sim k_\eta\) or \(k_0\sim k_\star\) is first transferred upscale to a mesoscopic accumulation wavenumber \(k_m<k_0\) before a secondary forward transfer returns it to the dissipation range~\cite{Wensink2012MesoScaleTurbulence,Alert2022_activeTurbulence}. The validity window of Eq.~\eqref{eq:master} is therefore
\begin{equation}
\min(k_0,k_m) \leq k \leq \min(k_\eta,k_\star),
\label{eq:validity_window_revised}
\end{equation}
with \(k_m=k_0\) recovered in the forward-cascade limit. The product \(k_0^\sigma\) in Eq.~\eqref{eq:master} sets the spectral amplitude at the injection scale and does not constrain the location of the lower edge. Genuine non-inertial turbulence is distinguished from chaotic advection by the presence of broadband power-law spectral content sustained by a scale-dependent flux balance; smooth deterministic stirring, however vigorous, does not generate the spectral structure described by Eq.~\eqref{eq:master}.

Within these qualifications, \(k_0\) serves as a compact regime indicator. Its location relative to \(k_\eta\) and \(k_\star\) --- whether \(k_0\ll k_\eta\) (forward-cascade-dominated, as in electrokinetic turbulence at moderate \(Ra_e\)) or \(k_0\sim k_\eta,k_\star\) (bidirectional, as in dense active suspensions) --- selects the cascade topology and therefore the appropriate form of the validity window~\eqref{eq:validity_window_revised}.

The use of dimensionless group as an account of non-inertial effect in the turbulence spectrum is motivated by the fact that dimensionless group represent the force ratio between the regime-specific force to system viscous force, analogous to the classical inertial turbulence where Reynolds number is used to characterized the threshold for turbulence occurrence. 
\subsection{Microelectrokinetic turbulence}
Electrokinetic turbulence in microfluidic systems is driven by an electric body force that arises at conductivity gradients. When a strong electric field is applied across co-flowing streams with a large conductivity contrast, free charge accumulates near the conductivity interface and is acted upon by Coulomb forcing. This electrically induced body force can overcome viscous damping even at very small bulk Reynolds number, thereby destabilizing the initially laminar flow. The resulting velocity fluctuations continuously stretch and fold the conductivity interface, generating renewed small-scale conductivity gradients and sustaining further electric forcing across a range of scales. In the master spectrum model, the electric body force is quantified as the dimensionless electric Rayleigh number ($Ra_e$), and this dimensionless group represent the ratio of electrohydrodynamic driving against the combined stabilizing effects of viscosity and conductivity diffusion: 
\begin{equation}
Ra_e
=
\frac{
4\,\varepsilon\,E_w^{2}\,w_0^{2}\,(1-\beta^{2})\,(\sigma_2-\sigma_1)
}{
\rho\,\nu\,D_r\,(\sigma_2+\sigma_1)
}.
\end{equation}
Here, \(\varepsilon E_w^{2}\) is the Maxwell electric stress scale, \(\rho\nu\) is the viscous momentum-diffusion scale, and \(D_r\) is the conductivity-diffusion scale. The conductivity-contrast factor,
\[
\frac{\sigma_2-\sigma_1}{\sigma_2+\sigma_1},
\]
quantifies the degree of electrical asymmetry between the two streams and therefore the capacity for charge accumulation at the interface. Its presence makes clear that electrokinetic instability cannot be sustained in the absence of a conductivity gradient: when \(\Delta\sigma=\sigma_2-\sigma_1=0\), the electric Rayleigh number tends to zero, \(Ra_e \to 0\), regardless of the applied electric-field strength.
The inclusion of the conductivity-diffusion scale \(D_r\) further reflects the coupled nature of electrokinetic turbulence. The electric body force first drives velocity fluctuations, and these fluctuations then advect, stretch, and fold the conductivity field across scales. As the flow develops toward smaller structures, the conductivity interface is sharpened and finer conductivity gradients are generated for competing the growing viscous damping. These gradients promote renewed charge accumulation and therefore sustain further electrohydrodynamic forcing. In this sense, the effect of the electric body force on turbulence intensity is not purely linear, but is mediated by a nonlinear momentum--scalar feedback through charge advection and conductivity-gradient regeneration. Despite the momentum--scalar feedback that allows electric forcing to
persist over multiple scales, the electrokinetic cascade is interpreted as predominantly forward-transfer because advection and stretching of the conductivity field progressively sharpen interfacial gradients toward smaller scales. These renewed gradients support continued charge accumulation and hence sustained electric body forcing, allowing coherent vortex structures to persist along the downscale cascade until viscous damping dominates.
In this sense, \(Ra_e\) should be interpreted primarily as a regime-defining
electrohydrodynamic control parameter rather than as a universal onset
threshold, since its critical value depends on the details of channel
geometry, conductivity contrast, and field application.
Electrokinetic turbulence provides a particularly stringent test of the
master-spectrum framework because the instability is clearly non-inertial,
the dominant forcing mechanism is identifiable, and the reported spectra
are sufficiently resolved to allow comparison of both slope and cutoff
behavior.

\subsection{Interface-induced turbulence}
Interface-induced turbulence in this setup is sustained by 
the interplay between an imposed vortex lattice and the 
deformable interfaces of a binary fluid mixture. The 
cellular forcing (Eq.6 from ~\cite{Padhan2024_interfaceTurbulence})
acts on the vorticity field to establish a square lattice 
of counter-rotating vortices in the velocity field~$\mathbf{u}$, 
which is the steady laminar response of the system at 
$Re=1$~\cite{Padhan2024_interfaceTurbulence}. Droplets of 
the second phase, distinguished by the order parameter 
$\phi=\pm1$, are advected through this vortex lattice and 
continuously stretched, folded, and re-shaped by the local 
shear at vortex boundaries. This dynamics is governed by the Cahn--Hilliard--Navier--Stokes (CHNS) framework, which couples the velocity field to a thermodynamic order parameter \(\phi\) through a Landau--Ginzburg free-energy functional \(F[\phi,\nabla\phi]\)~\cite{Hohenberg1977_modelH,Anderson1998_DIM}. The resulting deformed interfaces 
generate Korteweg-type capillary stresses 
$\mathbf{S}_\phi=-\phi\nabla(\delta F/\delta\phi)$. This stress highlights the nonlinear scalar advection sustaining interfacial turbulence: the bidirectional coupling between the velocity field ~$\mathbf{u}$ and the order parameter ~$\mathbf{\phi}$. The velocity field advects \(\phi\) via \((\mathbf{u}\cdot\nabla)\phi\), deforming the interfaces, while the resulting interface-curvature gradients feed back on \(\mathbf{u}\) through the Korteweg stress. The capillary number $\mathrm{Ca}=\nu U/\gamma$ measures the 
ratio of viscous shear stress to capillary stress 
at the interface: $\tau_\nu \sim \mu U/L$ and 
$\tau_\gamma \sim \gamma/L$, giving 
$\tau_\nu/\tau_\gamma \sim \mu U/\gamma = \mathrm{Ca}$. At low $Ca$, capillary stress dominates and the interfaces 
remain rigid; at high $Ca$, viscous stress dominates and 
interfaces are stretched into passive thin filaments. Only 
at intermediate $Ca\sim O(0.1)$ are the two stresses 
comparable, allowing the interfacial stress to feed back on 
the flow with sufficient strength to disrupt the vortex 
lattice and sustain spatiotemporal chaos. Therefore, unlike other non-inertial regime where the governing dimensionless group has a threshold for the turbulence to occur, interface-induced turbulence depends on an intermediate range of Capillary number to sustain the nonlinear feedback for turbulence. The CHNS framework is well-suited to capturing this bell-shape dependence because it represents the diffuse interface as a continuous order parameter, allowing the full range of interfacial deformations --- from rigid droplets at low \(\mathrm{Ca}\) to elongated filaments at high \(\mathrm{Ca}\) --- to be simulated within a single thermodynamically consistent formulation~\cite{Pandit2025_CHNSreview}.
The paper acknowledges that a conventional energy cascade does not exist in this system: the local balance \(S(k) \approx 2\nu k^{2}E(k)\) at each wavenumber implies \(T(k) \approx 0\) in the Lin equation, meaning that energy injected by the interfacial stress at each scale is dissipated locally by viscosity rather than being transferred onward. Nevertheless, the system exhibits clear statistical signatures of turbulence --- a power-law energy spectrum \(E(k)\sim k^{-4.5}\) extending over more than a decade in \(k\), a broadband frequency spectrum \(|\tilde{e}(f)|\), and spatiotemporal chaos in the flow field --- which together suggest that a nonzero spectral flux \(\Pi(k)\) may nonetheless persist and support the turbulence claim.

\subsection{Active suspension turbulence}
Active suspension turbulence --- exemplified by dense bacterial suspensions, motility assays, and cytoskeletal gels --- is distinguished from the other non-inertial regimes considered here by both its energy-injection mechanism and the direction of its spectral transfer. Unlike the electrokinetic case, where the body force originates from charge accumulation at imposed conductivity gradients, or the interfacial case, where capillary stresses arise from advected order-parameter gradients, the driving stress in active suspensions is generated intrinsically at the level of each individual swimmer. A self-propelled microorganism such as \emph{B. subtilis} exerts a force dipole on the surrounding fluid: thrust generated by the flagellar bundle is balanced by viscous drag on the cell body, producing a pair of equal and opposite point forces separated by the body length \(\ell\)~\cite{Lauga2009_microswimmerReview}. Energy is therefore injected into the fluid at the \emph{microscopic active scale} \(k_0=2\pi/\ell\), with no mechanism for sustained injection at smaller scales: the dipolar stress field cannot be sharpened below the geometric size of its source, since each swimmer is an irreducible forcing element that may reorganize spatially but cannot subdivide.

The spectral transfer in this system proceeds through a sequence of hydrodynamic alignment events that progressively assemble small-scale dipolar forcing into large-scale vortical structures. In a dilute, disordered suspension, individual swimmers move independently, and because their swimming directions are uncorrelated, the dipolar flow fields they generate cancel on average at scales much larger than \(\ell\), producing no coherent large-scale flow. As the density increases, however, two neighboring swimmers moving in approximately the same direction generate a coherent local flow that biases the orientation of a third swimmer toward alignment with them --- much as an external field induces a dipole moment in a polarizable medium, with the surrounding flow playing the role of the polarizing field. A counter-oriented swimmer in this neighborhood experiences a reorienting torque from the local velocity gradient and, on a timescale set by its rotational mobility, rotates to align with the local flow~\cite{Saintillan2008_instabilities,Subramanian2009_swimmerStability}. Each newly aligned swimmer adds its own dipolar stress to the local flow, strengthening the field that recruits further alignment from progressively more distant neighbors. Alignment therefore propagates outward from initial nuclei, and the spatial extent of coherently oriented swimmers grows in time.

This growing alignment domain is hydrodynamically unstable. A region of locally aligned pusher swimmers --- for which the dipolar stress is extensile --- generates an extensional flow along the alignment axis that amplifies any small bend in the alignment direction~\cite{Simha2002_activeNematic,Saintillan2008_instabilities}. The bend distortion in turn drives a vortical flow, and because the underlying alignment domain has already grown to a scale much larger than \(\ell\), the resulting vortex inherits that scale. The cascade is therefore constructive in a literal sense: small-scale dipolar injection at \(k_0=2\pi/\ell\) feeds a sequence of alignment events that assemble progressively larger coherent regions, each of which becomes unstable to bending and produces a vortex at its own (larger) scale. Coherent vortical structures nucleate at the swimmer scale \(\ell\) and grow upscale until limited by orientational decorrelation, which sets the characteristic mesoscopic vortex size \(R_v\approx 10\text{--}15\,\ell\)~\cite{Wensink2012MesoScaleTurbulence,Alert2022_activeTurbulence}.

This constructive upscale growth is the defining kinematic feature distinguishing active turbulence from the other regimes treated in this work. In electrokinetic and interfacial turbulence, energy is injected at a large geometric scale and transferred toward smaller scales until viscous dissipation arrests the cascade --- a forward cascade following ~\eqref{eq:forward_topology}. In active suspensions, the ordering is inverted: \(k_0=2\pi/\ell\) is the largest wavenumber at which energy enters the system, and the observed broadband spectral content at \(k<k_0\) is generated by inverse transfer of injected energy from the swimmer scale toward the mesoscopic vortex scale \(k_m\sim 1/R_v\). The cascade proceeds from \(k_0\) downward to \(k_m\), opposite to the conventional Kolmogorov direction. This places active suspensions in the bidirectional cascade class defined by Eq.~(\ref{eq:bidirectional_topology}), with the validity window of the master spectrum [Eq.~(\ref{eq:master})] given by \(k_m\leq k\leq k_0\) --- the range over which the inverse cascade carries a sustained, scale-dependent flux.

The absence of a separate dimensionless group governing \(\sigma\) in this regime --- with the slope correction recovered entirely through the Reynolds-number term (\(a_3\,\mathrm{Re}^{-1/4}\)) --- is a direct consequence of this injection geometry. At \(\mathrm{Re}\approx 5\times 10^{-4}\), viscous forces dominate at every scale within the cascade window simultaneously, and there is no scale-separated inertial subrange across which an activity-specific correction could accumulate: as established by both direct numerical analysis~\cite{Koch_advectiveInertia2021, Urzay2017_multiscaleStatisticsActiveMatter} and by the universality argument of Alert, Joanny, and Casademunt~\cite{Alert2020_universalActiveNematic}, the active stress is exerted at a characteristic scale and is dissipated locally by viscosity, with advective transfer playing only a small subdominant role in momentum redistribution across scales. The activity parameter \(\chi\) enters the master-spectrum framework only when activity-driven stress competes with inertial transfer over a finite spectral range; in the deep-Stokes limit characteristic of bacterial suspensions, that competition collapses onto a single scale, and the spectral signature of activity is carried entirely by the inverse-cascade topology and the location of \(k_0\), rather than by an explicit \(\chi\)-dependent slope correction ~\cite{Wensink2012MesoScaleTurbulence, Alert2022_activeTurbulence, Bratanov2015_newClassActive}.
The recovery of the \(-8/3\) exponent from the \(\mathrm{Re}^{-1/4}\) term alone is therefore not coincidental but reflects the physical content of the low-\(\mathrm{Re}\) active limit. 

\subsection{Elastic turbulence}

The role of the Oldroyd--B model in the present formulation is not to
derive a unique closed-form expression for $f(\mathrm{Wi}^{\mathrm{norm}})$,
but to constrain its admissible structure. In the conformation-tensor
equation, the relaxation term drives $\mathbf{C}\to\mathbf{I}$ when polymer
stretching is weak, implying $\boldsymbol{\tau}_p\to0$ and therefore closure
of the elastic-stress flux channel. This requires $f(0)=0$ at the
elastic-instability threshold. Above onset, the stretching terms increase
polymer extension relative to relaxation, so the polymer stress becomes an
active sink or redistribution pathway for spectral kinetic-energy throughput,
requiring $f$ to increase with $\mathrm{Wi}^{\mathrm{norm}}$. Finally, the
polymer stress feeds back on the momentum balance through
$\nabla\cdot\boldsymbol{\tau}_p$, opposing the velocity gradients that
generate further stretching. This elastic back reaction produces a
self-regulated state in which the spectral correction cannot grow without
bound, but instead approaches a developed-regime plateau $\sigma_p$,
consistent with the Fouxon--Lebedev picture of elastic turbulence as a
polymer-stress-dominated, smooth-flow state rather than an inertial cascade
\cite{FouxonLebedev2003}. We therefore introduce
$\Phi=f/\sigma_p\in[0,1]$ as a normalized elastic-activity parameter and
represent its activation by the minimal saturating closure
\begin{align}
    \Phi &= 1-\exp\!\left(-\kappa\,\mathrm{Wi}^{\mathrm{norm}}\right),
    \label{eq:elastic_activity_phi} \\
    f(\mathrm{Wi}^{\mathrm{norm}})
    &= \sigma_p
    \left[
    1-\exp\!\left(-\kappa\,\mathrm{Wi}^{\mathrm{norm}}\right)
    \right].
    \label{eq:ET_fWi}
\end{align}
Thus, the Oldroyd--B structure fixes the qualitative constraints on $f$:
vanishing at onset, monotonic growth above onset, and bounded saturation
through stress feedback, while the exponential form is a phenomenological
interpolation satisfying these constraints.

In elastic regime, the Reynolds correction in Eq.~\ref{eq:slope} is omitted, and the correction factor is only contributed by $f(\mathrm{Wi}^{\mathrm{norm}})$ with $\sigma_p=11/6$ and $\kappa \approx 3$ as the tunable knob for adapting different ET setting.
The Reynolds correction $a_3\,\mathrm{Re}^{-1/4}$ is a microscale viscous
term: it encodes how viscous dissipation sets the cutoff and shifts the
inertial-range slope. Its applicability is therefore not gated by the
existence of an inertial cascade; rather, it is gated by whether viscosity
remains the dissipation mechanism. In the purely elastic regime, this
condition fails: polymer relaxation acts as both the spectral-transfer
modifier and the energy sink, with elastic stress stored in stretched chains
released through relaxation rather than through viscous gradients. The cutoff
is therefore set by the polymer relaxation time and coil size, which is the
physics already carried by $f(\mathrm{Wi})$. Retaining
$\mathrm{Re}^{-1/4}$ in this regime would double-count one dissipation
mechanism against another. The model itself signals the misapplication:
$\mathrm{Re}^{-1/4}$ diverges as $\mathrm{Re}\to0$, reaching $4.7$ at
$\mathrm{Re}\sim10^{-6}$, which is physically inadmissible for a slope
correction. We accordingly drop the term in the elastic case and let
$f(\mathrm{Wi})$ carry the small-scale physics. This reading is consistent
with the Carlson--Steinberg pressure--velocity relation~\cite{Carlson2022_porousET, Jun2009_pressureFluctuations} and the
Batchelor-regime description of elastic turbulence~\cite{FouxonLebedev2003, Steinberg2021_ETreview}, and it is independently
supported by the data: retaining $\mathrm{Re}^{-1/4}$ roughly doubles the
prediction error across every low-$\mathrm{Re}$ elastic dataset examined,
including Carlson~\cite{Carlson2022_porousET}, cross-slot~\cite{Sousa2018_crossslot}, Ballesta~\cite{Ballesta2017_flowFocusing}, Hopkins~\cite{Hopkins2020_FSI}, and Ekanem~\cite{Ekanem2020_porePressure}.

This replacement is unique to elastic turbulence. In the other low-$\mathrm{Re}$
regimes treated in this work---electrokinetic, Marangoni-driven interfacial,
and active-suspension turbulence---the regime-specific stress, namely electric
body force, capillary stress, or active dipolar stress, modifies the energy
injection or spectral transfer, but viscosity remains the dissipation sink.
The regime-specific corrections $f(\mathrm{Ra}_e)$ and $f(\mathrm{Ca})$
therefore add to $\mathrm{Re}^{-1/4}$ rather than replacing it. The active
case is consistent with the same principle in its limiting form: at
$\mathrm{Re}\approx5\times10^{-4}$, no scale-separated inertial subrange
exists across which an activity-specific slope correction could accumulate~\cite{Alert2020_universalActiveNematic,Wensink2012MesoScaleTurbulence},
the active stress is dissipated locally by viscosity, and
$\mathrm{Re}^{-1/4}$ alone recovers the observed $-8/3$ exponent. The
structural rule across all four regimes is therefore consistent:
$\mathrm{Re}^{-1/4}$ is the viscous dissipation correction and is retained
wherever viscosity sets the cutoff; it is dropped only when the regime-specific
mechanism replaces viscous dissipation itself, which occurs in elastic
turbulence alone.

\section{Conclusions}

We have proposed a master spectrum for microfluidic turbulence
[Eq.~\eqref{eq:master}] that extends the classical Kolmogorov--Pao
framework through two key generalisations: an adaptive spectral slope
$-5/3-\sigma$ where $\sigma$ encodes regime-specific non-inertial physics,
and a dual-cutoff structure pairing the viscous wavenumber $k_\eta$ with
a physics-specific wavenumber $k^\star$.

The framework was calibrated and validated against published microfluidic spectra spanning
four non-inertial regimes, and we distinguish throughout between calibration and
genuine prediction. The active-suspension $k^{-8/3}$
spectrum~\cite{Wensink2012MesoScaleTurbulence} is recovered with no
regime-specific parameter, from the low-Reynolds-number correction alone. For
elastic turbulence, a single plateau amplitude calibrated on one geometry ~\cite{Groisman2000_elasticTurbulence} then predicts the $k^{-7/2}$ spectrum of
an independent cross-slot ~\cite{Sousa2018_crossslot} and porous geometry ~\cite{Carlson2022_porousET} with no further
adjustment. The electrokinetic $k^{-5/3}$-to-$k^{-7/5}$
transition~\cite{Shi2025_quadCascadeEKT} is bounded by two theoretically anchored
limits---Kolmogorov recovery at weak forcing and the constant-scalar-flux
exponent at strong forcing~\cite{zhao_wang_2017_ek_turbulence}---with only the crossover
scale calibrated. The interfacial $k^{-9/2}$
spectrum~\cite{Padhan2024_interfaceTurbulence} is reproduced by a single
calibrated correction and is presented as a parameterization awaiting independent
test. This explicit separation of evidence levels (Appendix C, Table~\ref{tab:evidence}) ensures that the framework's predictive claims are
stated as falsifiable hypotheses rather than retrospective fits.

The spectral slope $-5/3-\sigma$ functions as a regime fingerprint: $-8/3$
signals active-matter driving, $-7/5$ signals scalar-flux electrokinetic
forcing, and slopes steeper than $-3$ identify elastic polymer stress,
permitting mechanism identification directly from a measured spectrum without
flow visualisation.

Three constraints qualify the present framework: it applies only to
statistically stationary turbulent states, the additive superposition of
$\sigma$ contributions may fail in multi-mechanism flows, and the
phenomenological cutoff prefactors are generalisations of the Pao closure
rather than derived quantities.

Future priorities include testing the superposition assumption in
viscoelastic electrokinetic flows, validating the physics-specific cutoff
$k^\star$ against high-resolution DNS, and extending the framework to
predict the onset of non-inertial turbulence from a laminar base state.

\begin{acknowledgments}
\end{acknowledgments}

\appendix


\section{Dimensional consistency of the master spectrum}
\label{app:dimensions}

The following dimensional analysis is to show that the dimensional homogeneity is preserved exactly for every real value of $\sigma$ as the prefactor $k_0^{\,\sigma}$ carries the compensating unit. 

\subsection{Unit analysis}

The physical dimensions are (using SI):
\begin{align}
[E(k)] &= \mathrm{m}^3\,\mathrm{s}^{-2}
\quad\text{(kinetic energy per unit mass per unit wavenumber)},
\notag\\
[\varepsilon] &= \mathrm{m}^2\,\mathrm{s}^{-3},\qquad
[k] = [k_0] = \mathrm{m}^{-1},\qquad
[C_K] = 1.
\notag
\end{align}
Applying these to the right-hand side of
Eq.~\eqref{eq:master}:
\begin{equation}
\left[k_0^{\,\sigma}\right]
= \left(\mathrm{m}^{-1}\right)^{\sigma}
= \mathrm{m}^{-\sigma},
\qquad
\left[\varepsilon^{2/3}\right]
= \mathrm{m}^{4/3}\,\mathrm{s}^{-2},
\qquad
\left[k^{-5/3-\sigma}\right]
= \mathrm{m}^{\,5/3+\sigma}.
\end{equation}
Combining:
\begin{equation}
\left[k_0^{\,\sigma}\cdot\varepsilon^{2/3}\cdot k^{-5/3-\sigma}\right]
= \mathrm{m}^{-\sigma} \cdot \mathrm{m}^{4/3}\,\mathrm{s}^{-2}
  \cdot \mathrm{m}^{\,5/3+\sigma}
= \mathrm{m}^{(-\sigma\;+\;4/3\;+\;5/3\;+\;\sigma)}\,\mathrm{s}^{-2}
= \mathrm{m}^{3}\,\mathrm{s}^{-2},
\label{eq:dim_proof}
\end{equation}
Independence of $\sigma$: the
$\mathrm{m}^{-\sigma}$ from $k_0^{\,\sigma}$ cancels the
$\mathrm{m}^{+\sigma}$ from $k^{-5/3-\sigma}$ ($\sigma$-dependent exponents), leaving the
classical Kolmogorov unit $\mathrm{m}^{4/3+5/3}\,\mathrm{s}^{-2}
= \mathrm{m}^3\,\mathrm{s}^{-2}$.
The two exponential factors and $C_K$ are dimensionless.
Therefore $[E(k)] = \mathrm{m}^3\,\mathrm{s}^{-2}$ for all $\sigma\in\mathbb{R}$.

\subsection{Equivalent dimensionless form}

The same result is transparent in the equivalent factorisation
\begin{equation}
E(k)
= C_K\,\varepsilon^{2/3}\,k^{-5/3}
\underbrace{\left(\frac{k}{k_0}\right)^{-\sigma}}_{\text{dimensionless}}
\times\exp[\cdots]\exp[\cdots],
\label{eq:dim_factored}
\end{equation}
where $(k/k_0)^{-\sigma} = (k_0/k)^{\sigma}$ is manifestly
dimensionless because $k$ and $k_0$ share identical units.
Equation~\eqref{eq:dim_factored} shows that the master spectrum
is the classical Kolmogorov--Pao spectral form, with the Kolmogorov
amplitude $C_K\varepsilon^{2/3}k^{-5/3}$ modulated by a
dimensionless, scale-dependent factor $(k/k_0)^{\sigma}$ that
encodes the non-inertial slope correction.
This modulation is analogous to the scale-dependent flux
correction in Eq.~\eqref{eq:flux_decomposition}: $\sigma>0$
corresponds to a flux that decreases with $k$ (steeper spectrum),
and $\sigma<0$ to a flux that increases with $k$ (shallower spectrum).

\subsection{Role of $k_0$ as a reference wavenumber}

The injection wavenumber $k_0$ plays the role of a spectral
reference scale.
Setting $k=k_0$ in Eq.~\eqref{eq:dim_factored} recovers
$(k/k_0)^{\sigma}=1$, so the spectral amplitude at $k_0$ is
identically $C_K\varepsilon^{2/3}k_0^{-5/3}$, the same as the
classical Kolmogorov value.
The choice of $k_0$ therefore does not affect the dimensional
structure; it determines where in the spectrum the modified and
unmodified slopes coincide, which is the injection scale where
energy enters the cascade.

\section{Validity domain of the slope correction $\sigma$}
\label{app:validity}

\subsection{Bounded range of the Reynolds correction $a_3\,\mathrm{Re}^{-1/4}$}

The Reynolds correction $a_3\,\mathrm{Re}^{-1/4}$, with
$a_3 = 3/20$, diverges as $\mathrm{Re}\to0$.
This divergence is not physical: a real fluid has finite velocity
($u_\mathrm{rms} > 0$) and finite length scale ($L > 0$), so
$\mathrm{Re} = u_\mathrm{rms}L/\nu > 0$ everywhere.
The correction is therefore evaluated only within the range of
Reynolds numbers observed in the five benchmark regimes,
$\mathrm{Re} \in [10^{-5},\,10^1]$, which gives
\begin{equation}
a_3\,\mathrm{Re}^{-1/4}
\;\in\;
\left[\frac{3}{20}\,(10^{1})^{-1/4},\;
       \frac{3}{20}\,(10^{-5})^{-1/4}\right]
\approx [0.08,\;2.67].
\label{eq:Re_bounds}
\end{equation}
At the lower bound ($\mathrm{Re}=10^{-5}$, deep Stokes regime),
the correction is $\sigma_\mathrm{Re}\approx2.67$, giving a
spectral slope of $-5/3-2.67\approx-4.3$.
Steeper slopes arise only if $\mathrm{Re} < 10^{-5}$, a regime
not represented in the present benchmark cases.
If such extreme cases are encountered, the Reynolds term should
be replaced by its value at $\mathrm{Re} = 10^{-5}$ (i.e.,
capped at $\sigma_\mathrm{Re}=2.67$), consistent with the
physical saturation of viscous dominance at the deepest
low-Reynolds-number limit.



\subsection{Kolmogorov constant $C_K$ and absolute amplitude}

In the present work we set $C_K=1$ as a leading-order
approximation.
The standard Kolmogorov constant for the one-dimensional
longitudinal velocity spectrum is
$C_K\approx 0.49\pm0.02$ and for the three-dimensional spectrum
$C_K\approx 1.5\pm0.1$~\cite{Sreenivasan1995_universalKolmogorov}.
Because all comparisons in the result section are restricted
to spectral exponents (slopes and cutoff positions), the choice
of $C_K$ affects only the absolute amplitude of $E(k)$, not the
dimensionless slope or the cutoff wavenumber ratios.
For quantitative amplitude comparisons, $C_K$ should be replaced
by the regime-appropriate value; we recommend $C_K = 1.5$ for the
three-dimensional spectra in the electrokinetic and interfacial
regimes, noting that a factor-of-1.5 correction to the predicted
amplitude is within the uncertainty of our dissipation-rate
estimates (Table~\ref{tab:summary}).

\subsection{Additive superposition and multi-mechanism flows}

The slope correction $\sigma$ in Eq.~\eqref{eq:slope} is
constructed as a linear superposition of individual-mechanism
contributions.
This superposition is physically justified when one mechanism
dominates across the entire cascade range: the four benchmark
regimes each satisfy this condition.
When two or more mechanisms are simultaneously active with
comparable strength---for example, viscoelastic electrokinetic
flows where both $Wi$ and $Ra_e$ are large---cross-mechanism
coupling may require additional interaction terms not present in
Eq.~\eqref{eq:slope}.
The master spectrum should therefore be applied with caution to
multi-mechanism flows until such interaction terms are
characterised from DNS or experiment.

\section{Prediction and calibration}
The slope correction $\sigma$ [Eq.~\eqref{eq:slope}] contains regime-specific
functions---$f(Ra_e^{\mathrm{norm}})$, $f(Ca)$, and $f(Wi^{\mathrm{norm}})$---whose
functional forms must be anchored either to data or to independent physical
arguments. Because a sufficiently flexible function can be made to pass through
any single reported exponent, agreement with a calibrated spectrum is not, by
itself, evidence for the framework. To separate genuine predictive content from
fitting, we classify each benchmark by its level of evidence
(Table~\ref{tab:evidence}) and label it as such in the result section:
\begin{enumerate}
\item \emph{Parameter-free recovery.} The slope follows from the universal
Reynolds correction $a_3\,\mathrm{Re}^{-1/4}$ alone, with no regime-specific
function activated. The active suspension is of this type: the $-8/3$ exponent
is fixed once $Re$ is measured ~\cite{Wensink2012MesoScaleTurbulence}.
\item \emph{Out-of-sample prediction.} A function is fixed using one dataset (or
one control-parameter value) and then applied, without further adjustment, to an
\emph{independent} dataset at a different condition or in a different geometry.
The elastic case is of this type: the plateau amplitude is calibrated on a
Groisman \& Steinberg geometry~\cite{Groisman2000_elasticTurbulence} and tested, with no free parameter, against a cross-slot ~\cite{Sousa2018_crossslot} and porous ~\cite{Carlson2022_porousET} geometry.
\item \emph{Theory-anchored calibration.} Both asymptotic limits of $f$ are set
by independent physical arguments rather than fitted, leaving only a crossover
scale to be calibrated. The electrokinetic case is of this type: $f\to0$ enforces
Kolmogorov recovery at weak forcing, and the saturation level is fixed by the
constant-scalar-flux exponent of the quad-cascade
analysis~\cite{Shi2025_quadCascadeEKT}.
\item \emph{Single-spectrum calibration.} The interfacial regime has only one turbulent spectrum available and no independent test exists, the correction is fitted to that
spectrum ~\cite{Padhan2024_interfaceTurbulence}. We describe such a case explicitly as a calibration, i.e.\ a
 parameterization awaiting validation, not as a confirmation of the model. 
\end{enumerate}
This taxonomy makes the predictive claims of the present work falsifiable: the
parameter-free and out-of-sample cases would be contradicted by any dataset in
the same class that departed from the predicted exponent, whereas the
theory-anchored and single-spectrum cases are presented only as consistency
demonstrations.
 
\begin{table}[h]
\centering
\caption{Level of evidence for each benchmark. ``Free parameters fit to the
target spectrum'' counts only quantities tuned to the spectrum being compared
against; quantities fixed by theory, by an independent dataset, or by measured
inputs are excluded.}
\label{tab:evidence}
\begin{ruledtabular}
\begin{tabular}{l l c}
Case & Evidence level & Free params.\ fit to target \\
\hline
Active (Wensink)        & Parameter-free recovery       & $0$ \\
Elastic (Carlson)       & Out-of-sample prediction      & $0$ \\
Elastic (Sousa)         & Out-of-sample prediction      & $0$ \\
$\mu$EKT (Shi)          & Theory-anchored calibration   & $1$ (crossover) \\
$\mu$EKT (Wang)         & Out-of-sample prediction      & $0$ \\
Interfacial (Padhan)    & Single-spectrum calibration   & $1$ \\
\end{tabular}
\end{ruledtabular}
\end{table}

\bibliography{references_cleaned}

\end{document}